\documentclass[12pt]{article}
\usepackage{amsmath}
\usepackage[psamsfonts]{amssymb}

\date{\today}
\newcommand\email[1]{{\tt {#1}}}
\def\acknowledgments{\centerline{{\bf Acknowldegments}}}
\def\disclaimer{\centerline{{\bf Disclaimer}}}
\begin{document}
\setlength{\baselineskip}{20pt}

\begin{titlepage}
\rightline{hep-th/0311042}
\rightline{UCB-PTH-03/30}
\rightline{LBNL-53980}
\vskip 1cm
\begin{center}
\Huge
Twisted Six Dimensional Gauge Theories on Tori,\\
        Matrix Models, and Integrable Systems \\
\end{center}
\vskip 0.5cm\centerline
{
Surya Ganguli, Ori J. Ganor and James Gill}
\begin{center}
\em Department of Physics\\
\em University of California, \\ and\\
\em Theoretical Physics Group \\
\em Lawrence Berkeley National Laboratory\\
\em Berkeley, CA 94720\\
Email: \email{sganguli, origa, gill@socrates.berkeley.edu}
\end{center}


\abstract{ We use the Dijkgraaf-Vafa technique to study massive vacua
of 6D SU(N) SYM theories on tori with R-symmetry twists. One finds a
matrix model living on the compactification torus with a genus 2
spectral curve.  The Jacobian of this curve is closely related to a
twisted four torus $T$ in which the Seiberg-Witten curves of the
theory are embedded.  We also analyze R-symmetry twists in a bundle
with nontrivial first Chern class which yields intrinsically 6D SUSY
breaking and a novel matrix integral whose eigenvalues float in a
sea of background charge.  Next we analyze the underlying integrable
system of the theory, whose phase space we show to be a system of N-1
points on $T$.  We write down an explicit set of Poisson commuting
Hamiltonians for this system for arbitrary N and use them to prove
that equilibrium configurations with respect to all Hamiltonians
correspond to points in moduli space where the Seiberg-Witten curve
maximally degenerates to genus 2, thereby recovering the matrix model
spectral curve.  We also write down a conjecture for a dual set of
Poisson commuting variables which could shed light on a particle-like
interpretation of the system. }
\end{titlepage}

\def\be{\begin{equation}} 
\def\ee{\end{equation}} 
\def\bear{\begin{eqnarray}} 
\def\eear{\end{eqnarray}} 
\def\nn{\nonumber}

\def\defineas{{\stackrel{{\mbox{\tiny def}}}{=}}} 
 
\def\a{\alpha} 
\def\b{\beta} 
\def\g{\gamma} 
\def\u{\mu} 
\def\v{\nu} 
\def\r{\rho} 
\def\th{{\theta}} 
\def\lam{{\lambda}} 

\def\bth{{\overline{\theta}}} 
\def\blam{{\overline{\lambda}}} 
\def\bpsi{{\overline{\psi}}} 
\def\bsig{{\overline{\sigma}}} 

\def\rt{{\rightarrow}}  
\def\cc{{\mbox{c.c.}}} 

\def\hI{{\hat{I}}} 
\def\ha{{\hat{a}}} 
\def\hadg{{\hat{a}^\dagger}} 
\def\hPhi{{\hat{\Phi}}} 
\def\btau{{\overline{\tau}}} 

\newcommand\SUSY[1]{{${\cal N}={#1}$}}  
\newcommand\px[1]{{\partial_{#1}}} 
\newcommand\qx[1]{{\partial^{#1}}} 
\newcommand\Del[1]{{D_{#1}}} 

\newcommand\bra[1]{{\langle {#1}|}} 
\newcommand\ket[1]{{|{#1}\rangle}} 

\def\Z{{\mathbb Z}}
\def\R{{\mathbb R}}
\def\C{{\mathbb C}}
\def\Id{{\mathbb I}}

\def\nwH{{\widetilde{H}}}
\def\nwM{{\widetilde{M}}}
\def\tr{{\mbox{tr}\,}}
\def\Pv{{\cal P}} 
\def\bz{{\overline{z}}}
\def\gYM{{g_{\mbox{\rm\tiny YM}}}} 
\def\gST{{g_{\mbox{\rm\tiny st}}}} 
\newcommand\rep[1]{{\bf {#1}}} 
\def\tw{{\alpha}} 

\def\cV{{\cal V}} 
\def\cA{{\cal A}} 
\def\cP{{\Phi}} 
\def\cZ{{\cal Z}} 

\newcommand\wl[1]{{\lambda_{#1}}} 
\def\hWl{{\hat{W}}} 
\def\LBd{{\cal L}} 
\def\Scn{{\sigma}} 
\def\mod{{\mbox{\rm mod}\,}} 
\def\vomega{{\vec{\omega}}} 

\def\vth{{\vartheta}} 
\newcommand\thx[2]{{\vartheta
   \left[{#1}\atop{#2}\right]}} 
\newcommand\thxx[4]{{\vartheta\left[
      {{#1}\atop{#2}}\atop{{#3}\atop{#4}}\right]}} 

\def\rB{{A^{({\mbox{\rm\tiny n.d.}})}}} 

\newcommand\Jac[1]{{\ensuremath{{\mathcal{J}}({#1})}}}  
\newcommand\Symm[1]{{\ensuremath{\mathrm{Symm}{#1}}}}  
\def\Msw{{\cal M_{\mathrm{SW}}}}
\def\P{{\mathbb P}}

\section{Introduction}\label{sec:intro}
Recent advances in the study of \SUSY{1} supersymmetric gauge theories
\cite{Dijkgraaf:2002fc}-\cite{Cachazo:2002ry}
have provided \cite{Cachazo:2002zk}-\cite{Alishahiha:2003pu}
a unifying framework for deriving exact results
for superpotentials of \SUSY{1} theories
\cite{Veneziano:1982ah}-\cite{Seiberg:1995ac}
and moduli spaces of \SUSY{2} theories
\cite{Seiberg:1994rs}-\cite{Seiberg:1994aj}.

One of the remarkable aspects of the Dijkgraaf-Vafa (DV) technique
\cite{Dijkgraaf:2002fc} is that it can also be applied to
nonrenormalizable \SUSY{1} theories. For such theories, there still
exists a subset of questions with finite answers that are independent
of the UV regularization.  These are the questions pertaining to the
chiral ring and the DV technique provides a solution to those
questions.

This observation allows one to study higher dimensional gauge theories
\cite{Dijkgraaf:2003xk}-\cite{Hollowood:2003gr}.  The purpose of this
paper is to explore the application of the DV technique to 5+1D
Super-Yang-Mills theories and to relate this approach to the viewpoint
of integrable systems. The DV technique applies most directly to
3+1D gauge theories with \SUSY{2} or \SUSY{1} supersymmetry. We will
therefore compactify the 5+1D gauge theories on $T^2$ to 3+1D. A
simple way to break supersymmetry to \SUSY{2} is to introduce
R-symmetry twists. The 5+1D theory has $SU(2)\times SU(2)$ R-symmetry
which can be broken down to $SU(2)\times U(1)$ with appropriate
boundary conditions (i.e. R-symmetry Wilson lines).

The resulting 3+1D low-energy effective action was studied in
\cite{Cheung:1998te}-\cite{Cheung:1998wj}. The effective action is
described in terms of Seiberg-Witten curves that can be embedded inside
a fixed $T^4$.  The complex structure of the $T^4$ should be
determined from the complex structure of the physical $T^2$, the ratio
between its area and the 5+1D Yang-Mills coupling constant squared,
and the values of the R-symmetry twists.

One can also compactify on $T^2$ with more complicated R-symmetry
boundary conditions so as to preserve only \SUSY{1} in 3+1D.  One way
to do that was studied in \cite{Chan:2000qc}.  For this purpose a
$U(1)$ subgroup of the R-symmetry group is picked and then twisted so
as to form a $U(1)$ bundle with nonzero first Chern class over $T^2$.

In this paper we will study these resulting 3+1D theories from the
perspective of the DV conjecture and integrable systems. The paper is
organized as follows.  In section (\ref{sec:setup}) we explain the
setup starting from 5+1D and compactifying down to 3+1D. In section
(\ref{sec:DVSW}) we write down and solve the matrix model required to
obtain the degenerated genus 2 Seiberg-Witten curves of the
compactified theories in their massive vacua.  In section
(\ref{sec:DVSP}) we show how putting R-symmetry twists in a bundle of
nontrivial Chern class over the compactification torus yields a
modified matrix integral that could explain an intrinsically six
dimensional mechanism for breaking SUSY to \SUSY{1}.  We move on to
the integrable systems approach in section (\ref{sec:intsys}).  Here we review
the various relations between extremization of DV superpotentials,
degenerating Seiberg-Witten curves, and equilibrium configurations of
integrable systems.  We then use algebraic-geometry techniques (a sort
of Fourier-Mukai transform) to reformulate the integrable system
associated to the 6D theory in terms of a set of points on $T^4$.
This technique also yields an explicit set of Poission commuting
Hamiltonians for this system, written in terms of theta functions on
$T^4$.  We prove that configurations that are at equilibrium with
respect to all such Hamiltonian flows imply a degeneration of the
Seiberg-Witten curve to genus 2.  In section \ref{sec:DualHams} we
write down a conjecture for an alternate set of Poisson commuting
observables that could yield physical insight into the nature of the
integrable system.  We end in section \ref{sec:discussion} with
conclusions and directions for future work.  Also, for the convenience
of the reader we assemble facts about elliptic functions on $T^2$ in 
appendix \ref{sec:elliptic} and in appendix \ref{sec:ThetaApx} we
review higher dimensional theta functions, Jacobians of Riemann
surfaces and the Abel-Jacobi map.

\par Some of the earlier results in this paper appeared independently in 
\cite{Hollowood:2003cv}, where the matrix model was used to extract 
Seiberg-Witten curves.  Also, upon completion of this paper, \cite{Braden:2003} 
appeared which also discusses the integrable systems viewpoint.

\section{The setup: 5+1D gauge theories on $T^2$}\label{sec:setup}
Our starting point is 5+1D Super-Yang-Mills theory with gauge group
$SU(N)$ and with \SUSY{(1,1)} supersymmetry.
The coupling constant $\gYM$ has dimensions of length.
We denote the coordinates by $x_0\dots x_5$ and we set
$$
z\defineas x_4 + i x_5,\qquad \bz\defineas x_4 -i x_5.
$$
The R-symmetry is $Spin(4)= SU(2)\times SU(2)$.
Our indexing notation is as follows
\vskip 0.5cm
\begin{tabular}{lll}\hline
Symbol & Range & Representation \\ \hline\hline
$\u,\v,\cdots$ & $0\dots 5$ & Spacetime vector \\
$a,b,\cdots$ & $1,2$ & R-symmetry ``left'' spinor \\
$\dot{a},\dot{b},\cdots$ & $\dot{1},\dot{2}$ & R-symmetry ``right'' spinor \\
$\a,\b,\cdots$ & $1,\dots,4$ & spacetime left Weyl spinor \\
$\dot{\a},\dot{\b},\cdots$ & $\dot{1},\dots,\dot{4}$ & spacetime right Weyl spinor \\
\hline
\end{tabular}
\vskip 0.5cm
The field contents of the theory is described in the following table.
\vskip 0.5cm
\begin{tabular}{llc}\hline
Symbol & Field & $SU(2)\times SU(2)$ representation \\ \hline\hline
$A_\u$ & Gauge & $(\rep{1}, \rep{1})$ \\
$\Phi^{\a\dot{\b}}$ & Scalar & $(\rep{2}, \rep{2})$ \\
$\psi^{a\a}$ & Left moving fermion & $(\rep{2},\rep{1})$ \\
$\psi^{\dot{a}\dot{\a}}$ & Right moving fermion & $(\rep{1},\rep{2})$ \\ \hline
\end{tabular}
\vskip 0.5cm
The Lagrangian is 
\bear
{\cal L} &=& \frac{1}{\gYM^2}\tr\left\{\frac{1}{4}F_{\u\v}F^{\u\v}
+ \frac{1}{2}D_\u\Phi_{\a\dot{\a}} D^\u\Phi^{\b\dot{\b}}
-\frac{1}{2}[\Phi^{\a\dot{\a}},\Phi^{\b\dot{\b}}][\Phi_{\a\dot{\a}},\Phi_{\b\dot{\b}}]
   + {\mbox{fermions}}\right\}.
\nn
\eear

\subsection{Compactification on $T^2$ with twists}\label{subsec:tw}
\vskip 0.5cm
\begin{figure}[t]
\begin{picture}(400,130)

\thinlines
\multiput(10,10)(50,100){2}{\line(1,0){100}} %
\multiput(10,10)(100,0){2}{\line(1,2){50}} %
\put(75,55){$z$}
\put(60,100){$\tau$}
\put(10,0){\vector(1,0){100}} \put(115,0){$e^{i\tw_1}$}
\put(0,10){\vector(1,2){50}} \put(50,115){$e^{i\tw_2}$}
\end{picture}
\caption{
The boundary conditions along the $T^2$ in the compact directions $4-5.$
The scalar and fermion fields $\Phi^{\a\dot{\a}}$ and 
$\psi^{\dot{a}\dot{\a}}$ pick up phases when translated along cycles of $T^2.$}
\end{figure}
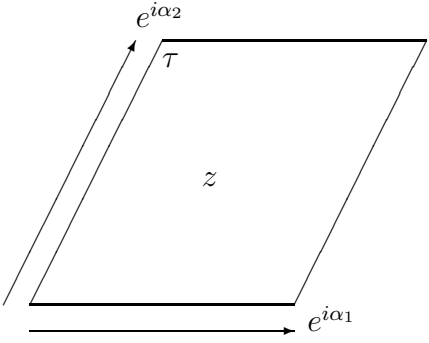
Let us now compactify on a torus $\Sigma_\tau$ with complex structure $\tau$ and
area $A$.
We choose the complex coordinate $z$ such that
$$
z\sim z+1\sim z+\tau.
$$
We can use one of the two $SU(2)$ factors of the R-symmetry to twist
the compactification. For this purpose we choose a $U(1)\subset SU(2)$
of the second factor (corresponding to the dotted $\dot{\a},\dot{\b}$ indices).
We then pick two constant elements in this $U(1)$ subgroup 
which we represent as $e^{i\tw_1}, e^{i\tw_2}$.
Here $\tw_1,\tw_2$ are constant phases corresponding to boundary conditions
in the $z\sim z+1$ and $z\sim z+\tau$ directions, respectively.
Explicitly, we set the boundary conditions
\bear
\Phi^{\a\dot{1}}(z) &=& e^{i\tw_1}\Phi^{\a\dot{1}}(z+1) 
      =e^{i\tw_2}\Phi^{\a\dot{1}}(z+\tau),\qquad
\Phi^{\a\dot{2}}(z) =e^{-i\tw_1}\Phi^{\a\dot{2}}(z+1) 
      =e^{-i\tw_2}\Phi^{\a\dot{2}}(z+\tau),
\nn\\
\psi^{\dot{a}\dot{1}}(z) &=& e^{i\tw_1}\psi^{\dot{a}\dot{1}}(z+1) 
      =e^{i\tw_2}\psi^{\dot{a}\dot{1}}(z+\tau),\qquad
\psi^{\dot{a}\dot{2}}(z) =e^{-i\tw_1}\psi^{\dot{a}\dot{2}}(z+1) 
      =e^{-i\tw_2}\psi^{\dot{a}\dot{2}}(z+\tau).
\nn\\
\label{twbc}
\eear
The remaining fields have periodic boundary conditions.

For nonzero phases $\tw_1,\tw_2$ this compactification breaks half of
the supersymmetry and preserves \SUSY{2} in the remaining noncompact 3+1D.
The fields in (\ref{twbc}) become massive and classically we are left with a 
massless  \SUSY{2} vector multiplet in 3+1D.

\subsection{Compactification on $T^2$ with with nonzero $c_1$}
\label{subsec:chern}
We can break the supersymmetry down to \SUSY{1} in 3+1D by introducing
a nonzero Chern class for the $U(1)$-bundle over $T^2$ \cite{Chan:2000qc}.
As before we fix a $U(1)\subset SU(2)$ subgroup of the second $SU(2)$ factor
of the R-symmetry group. We then pick a nondynamical $U(1)$ gauge field $\rB$
with components $\rB_z(z,\bz), \rB_\bz(z,\bz)$ only along $T^2$ and
with constant field-strength such that $\int_{T^2} d\rB = c_1 = n.$

We then take periodic boundary conditions for all the fields
but modify the covariant derivatives of the fields that are charged
under $U(1)$ 
[i.e. the fields appearing in (\ref{twbc})] to include the nondynamical gauge
field $\rB$.

As explained in \cite{Chan:2000qc}, such a modification to the Lagrangian
affects the fermions and scalars differently but it is possible to preserve
partial supersymmetry by adding an explicit coupling to one of the components
of the D-terms of the theory. The extra coupling is of the form

$$
\int \px{[z}\rB_{\bz]} D^{(3)}\, d^2z,\qquad
D^{(3)} \equiv 
  \Phi^{1\dot{1}}\Phi^{2\dot{2}} +\Phi^{1\dot{2}}\Phi^{2\dot{1}}.
$$
Here $D^{(3)}$ is a member of an $SU(2)_R$ triplet.
This term explicitly breaks the R-symmetry down to $U(1)$, and the
unbroken supersymmetry is indeed just \SUSY{1}.

\subsection{Quiver theories}
We can generalize the discussion above to quiver theories.
These are chiral gauge theories with \SUSY{(1,0)} in 5+1D.
The gauge group is 
$$
G = SU(N)_1\times SU(N)_2\times\cdots\times SU(N)_k
$$
where the subscript is a label of the factor.
The R-symmetry is $SU(2)_R$ and there is an additional global
$U(1)$ flavor symmetry.
The fields of the theory fall into vector multiplet and
hypermultiplet representations of \SUSY{(1,0)} supersymmetry.
The vector multiplet is in the adjoint representation of the gauge group
and its field content is described in the following table.
\vskip 0.5cm
\begin{tabular}{llcc}\hline
Symbol & Field & $SU(2)_R$ representation & $U(1)$ charge\\ \hline\hline
$A_\u$ & Gauge & $\rep{1}$ & $0$ \\
$\chi^{a\a}$ & Left moving gluino & $\rep{2}$ & $0$ \\ \hline
\end{tabular}
\vskip 0.5cm
There are $k$ hypermultiplets. The $j^{th}$ ($j=1\dots k$)
hypermultiplet has fields in the product of the 
anti-fundamental representation 
$\overline{N}$ of $SU(N)_j$ and the fundamental representation $N$
of $S(N)_{j+1}$ (with the cyclic convention $k+1\equiv 1$).
Its field contents is described in the following table.
\vskip 0.5cm
\begin{tabular}{llcc}\hline
Symbol & Field & $SU(2)_R$ representation & $U(1)$ charge\\ \hline\hline
$\Phi^{\a}_{j,j+1}$ & complex scalar & $\rep{2}$ & $+1$ \\
$\psi^{\dot{a}}_{j,j+1}$ & right moving spinor & $\rep{1}$ & $+1$ \\ \hline
\end{tabular}
\vskip 0.5cm
We can proceed to compactify these theories on $T^2$ either with \SUSY{2}
preserving $U(1)$ twists as in (\ref{subsec:tw}) or with a \SUSY{1} preserving
nontrivial $U(1)$ bundle with first Chern class $c_1 = n$ as in (\ref{subsec:chern}).
In the latter case, we obtain $n$ generations of chiral matter in 3+1D
\cite{Chan:2000qc}.

\section{\SUSY{1} Massive vacua from DV matrix integrals}\label{sec:DVSW}

We will now use the DV technique to probe the massive vacua of
\SUSY{1} deformations of the twisted 3+1D \SUSY{2} theory described in
section (\ref{subsec:tw}).  We use the same notation as in section
(\ref{subsec:tw}) but for convenience take the area $A$ of $\Sigma_\tau$
to be 1, which just sets the effective bare 3+1D gauge coupling to be
the same as the 5+1D coupling.  The starting point for implementing
the DV technique is to write the Lagrangian of the 5+1D theory in 3+1D
\SUSY{1} superspace.  The twists $e^{i\tw_1}, e^{i\tw_2}$ are in a
global $U(1)$ group.  The 5+1D vector field decomposes into a 3+1D
vector field, that is part of a vector multiplet $\cV$, and a 3+1D
complex scalar field $A_\bz\equiv A_4 - i A_5$ that is the scalar
component of a chiral multiplet $\cA_\bz$. The 5+1D scalars fall into
two 3+1D chiral multiplets $\cP_{+}, \cP_{-}$ with opposite $U(1)$
charges.  All the fields are functions of $z,\bz$ and are summarized,
together with their global $U(1)$ charges, in the following table:
\vskip 0.5cm
\begin{tabular}{lllc}\hline
Symbol & 5+1D Field & Multiplet & $U(1)$ charge \\ \hline\hline
$\cV$ & Gauge & vector & $0$ \\
$\cA_\bz$ & Gauge & chiral & $0$ \\
$\cP_{+}$ & Hyper & chiral & $+1$ \\
$\cP_{-}$ & Hyper & chiral & $-1$ \\ \hline
\end{tabular}
\vskip 0.5cm
The DV technique states that the properties of the chiral ring
of the effective 3+1D theory above can be deduced from the path integral of
the internal two dimensional ``holomorphic'' bosonic gauge theory,
\be\label{DVmm}
\cZ = \int [D\cA_\bz][D\cP_{+}][D\cP_{-}] e^{-\frac{1}{\gST^2}W}
\ee
where the superpotential is 
\be\label{WcPcP}
W = \int_{\Sigma_\tau} d^2z \tr\{\cP_{+}\px{z}\cP_{-} -i\cP_{+}[\cA_\bz,\cP_{-}] + W[\cA_\bz]\}.
\ee
Here $W[A_\bz]$, which we insert by hand to break supersymmetry
down to \SUSY{1}, can be an arbitrary gauge invariant 
holomorphic function of the chiral field
$\cA_\bz$. This superpotential is the dimensional reduction down to 2D
of the holomorphic Chern-Simons action in 6D that appears when
the DV technique is applied to 10D Super-Yang-Mills theory \cite{Dijkgraaf:2003xk}.
All fields are $\hat N \times \hat N$ matrices where $\hat N$ has no relation to the
physical $N$ of $SU(N)$. The fields $\cP_{+}$ and $\cP_{-}$ 
are required to have boundary conditions
\be\label{ptwbc}
\cP_{\pm}(z) = e^{\pm i\alpha_1}\cP_{\pm}(z+1) = e^{\pm i\alpha_2}\cP_{\pm}(z+\tau).
\ee
We must now specify the path in field space over which the
path integral in (\ref{DVmm}) should be performed.
Recall that in the 3+1D DV technique the path integral is of the form
$$
\int [D\cP] e^{-\frac{1}{g^2}W(\cP)}
\longrightarrow
\int\prod_{j=1}^{\hat N} d\lam_i\,\prod_{i<j}(\lam_i-\lam_j)^2
e^{-\frac{1}{g^2}\sum_i W(\lam_i)}.
$$ where $\lam_1,\dots,\lam_N$ are the complex eigenvalues of $\Phi$.
Note that the integral is performed only over real $\lam_i$'s (or any
other suitably chosen path in the complex $\lam_i$-plane).  In other
words, the measure is $\prod_i d\lam_i$ and not $\prod_i d^2\lam_i.$
Thus, the integration over $[D\cP]$ is performed not over the entire
$\cP$-space but only over the subspace restricted by $\cP^\dagger =
\cP$.  Similarly in the 5+1D case, one can choose the real slice
$\cP_+^\dagger = \cP_-$.  However in the integral over the
antiholomorphic connection $\cA_{\bz}$, the only gauge invariant data to be
integrated over are the holonomies $\hWl_1$ and $\hWl_2$ of $\cA_{\bz}$
around the two one cycles of $\Sigma_\tau$.  These are defined as

$$
\hWl_1 \defineas e^{i\oint_{z\rightarrow z+1} \cA_\bz d\bz},\qquad
\hWl_2 \defineas e^{i\oint_{z\rightarrow z+\tau} \cA_\bz d\bz},\qquad
\hWl_1, \hWl_2\in U(N),\qquad \hWl_1 \hWl_2 = \hWl_2 \hWl_1.
$$
By gauge transformations, we can simultaneously diagonalize $\hWl_1, \hWl_2$

\be\label{WlWl}
\hWl_1 =
\left(\begin{array}{cccc}
e^{2\pi i\lam^{(1)}_1} & & & \\ & e^{2\pi i\lam^{(2)}_1} & & \\
& & \ddots & \\ & & & e^{2\pi i\lam^{(N)}_1} \\
\end{array}\right),\qquad
\hWl_2 =
\left(\begin{array}{cccc}
e^{2\pi i\lam^{(1)}_2} & & & \\ & e^{2\pi i\lam^{(2)}_2} & & \\
& & \ddots & \\ & & & e^{2\pi i\lam^{(N)}_2} \\
\end{array}\right)\\,
\ee

\noindent and then combine the two sets of eigenvalues into one set of complex variables
\be\label{wlj}
\wl{j} \defineas \lam^{(j)}_2 + \lam^{(j)}_1\tau,\qquad j=1\dots \hat N.
\ee
The periodicity of the phases $\lam^{(j)}_1, \lam^{(j)}_2$ implies
that $\wl{j}$ is naturally defined on a (dual) $T^2$ of complex structure $\tau$,
$$
\wl{j}\sim \wl{j} + 1\sim \wl{j}+\tau.
$$ 

\par So the path integral over $\cA_\bz$ reduces to a finite dimensional
integral over $\lam_1,..\lam_{\hat N}$, but the change of variables incurs a
nontrivial Jacobian factor.  Since each integration variable $\lam_i$
naturally takes values on a torus, one can view this finite dimensional
integral simply as a compactified Hermitian matrix integral.  As is
well known in compactifying Matrix theory on tori, one thinks of each
eigenvalue $\lam_i$ as living on the complex plane; then adding in all images
$\lam_i + m + n\tau$, with $m,n \in \Z$, effectively compactifies
this plane.  The Jacobian of the change of measure is then simply
\be\label{DVJac}
\prod_{i<j}
  \prod_{m,n}(\wl{i}-\wl{j} + m + n\tau)^2
\ee
which is similar to the Vandermonde determinant appearing in the usual
Hermitian matrix integrals, but with contributions from differences
of eigenvalue images not related by $z\sim z+1\sim z+\tau$.
Furthermore, gauge invariance dictates that
our added superpotential $W$ is an elliptic function
in $\lam_i$.  

\par
Upon performing the change of variables (\ref{WlWl}) and (\ref{wlj}),
and taking into the account the Jacobian factor (\ref{DVJac}), the path
integral (\ref{DVmm}) becomes
$$
\cZ = \int d\wl{1}\dots d\wl{\hat N} \prod_{i<j}
  \prod_{m,n}(\wl{i}-\wl{j} + m + n\tau)^2
  [D\Phi_+][D\Phi_-]\exp{[-\frac{1}{\gST^2}\int_{T^2}
  \tr\Phi_-\Del{\bz}\Phi_+ + \sum_i W(\wl{i})] }.
$$
The integral over $[D\Phi_+][D\Phi_-]$ is gaussian and reduces to
calculating the functional determinant $\det[\Del{\bz}]$.  This can be
done by explicitly calculating the eigenvalues of $\partial_\bz$ in the space
of functions on the dual $\Sigma_\tau$ obeying specific boundary conditions.
For the $ij$th element of $\Phi_+$ these boundary conditions correspond
to picking up a phase $\exp{i[\lam^{(i)}_1 - \lam^{(j)}_1 + \alpha_1]}$ under
$ z \rightarrow z+1$ and $\exp{i[\lam^{(i)}_2 - \lam^{(j)}_2 +
    \alpha_2]}$ under
$ z \rightarrow z+\tau$.  In this space of functions the product of eigenvalues
of $\partial_\bz$ can be written as
$$
  \prod_{m,n}(\lam_{i}-\lam_{j} + \alpha + m + n\tau), \qquad m,n \in \Z.
$$
Here we have defined the new complex twist parameter
$$
\alpha \equiv \alpha_2 + \alpha_1\tau,
$$
and the $\lam_i$ are defined in \eqref{wlj}.  Thus integrating out
$\Phi_+$,
$\Phi_-$ leaves us with the integral

\bear \cZ = \int d\wl{1}\dots d\wl{\hat N} \prod_{i<j}
\frac{\prod_{m,n}(\wl{i}-\wl{j} + m + n\tau)^2}
{\prod_{m,n}(\wl{i}-\wl{j} + \tw + m + n\tau) (\wl{i}-\wl{j} - \tw + m +
n\tau)} e^{-\frac{1}{\gST^2}\sum_i W(\wl{i})} \nn \eear 

\begin{figure}[t]
\begin{picture}(520,220)
\put(0,0){\begin{picture}(150,150)
\put(85,120){(a)}
\thicklines
\put(0,10){\vector(1,2){25}} 
\put(25,60){\line(1,2){25}} 
\put(13,66){$B_1$}
\put(80,10){\line(1,2){50}} %
\put(50,110){\line(1,0){80}} %
\multiput(30,30)(7,0){6}{\line(1,0){5}} 
\multiput(60,90)(7,0){6}{\line(1,0){5}} 
\put(0,10){\vector(1,0){40}} 
\put(40,10){\line(1,0){40}} 
\put(35,0){$A_1$}
\thinlines
\put(80,90){\vector(-1,-2){30}} 
\put(50,30){\vector(1,2){30}} 
\put(45,112){$\tau$} 
\put(110,100){$z$} 
\put(80,0){$1$} 
\put(80,90){\begin{picture}(60,20)
\qbezier(30,0)(30,2)(28,4)
\qbezier(30,0)(30,-2)(28,-4)
\qbezier(-30,0)(-30,-2)(-28,-4)
\qbezier(-30,0)(-30,2)(-28,4)
\qbezier(28,4)(25,6)(21,7)
\qbezier(28,-4)(25,-6)(21,-7)
\qbezier(-28,-4)(-25,-6)(-21,-7)
\qbezier(-28,4)(-25,6)(-21,7)
\qbezier(21,7)(17,8)(11,9)
\qbezier(21,-7)(17,-8)(11,-9)
\qbezier(-21,-7)(-17,-8)(-11,-9)
\qbezier(-21,7)(-17,8)(-11,9)
\qbezier(11,9)(6,10)(0,10)
\qbezier(11,-9)(6,-10)(0,-10)
\qbezier(-11,-9)(-6,-10)(0,-10)
\qbezier(-11,9)(-6,10)(0,10)
\end{picture}} 
\put(80,100){\vector(1,0){0}} 
\put(90,70){$A_2$} 
\put(50,60){$B_2$} 
\put(75,58){$\alpha$} 
\end{picture}}

\put(120,0){\begin{picture}(100,60)
\thinlines
\put(0,40){\line(-1,1){10}} 
\put(0,60){\line(-1,-1){10}} 
\put(50,50){\line(-1,0){60}} 
\put(10,55){$z=u_1$} 
\end{picture}}

\put(180,0){\begin{picture}(150,150)
\put(50,120){(b)}
\thinlines
\put(0,10){\begin{picture}(70,70)
\multiput(0,0)(0,70){2}{\line(1,0){70}}
\multiput(0,0)(70,0){2}{\line(0,1){70}}
\end{picture}} 
\put(0,80){\line(1,1){30}} 
\put(70,80){\line(1,1){30}} 
\put(70,10){\line(1,1){30}} 
\put(100,40){\line(0,1){70}} 
\put(30,110){\line(1,0){70}} 
\put(40,30){$\Sigma_{\tau,\alpha,\tilde{g}}$} 
\thicklines
\qbezier(60,20)(10,20)(30,40) 
\qbezier(30,40)(50,60)(10,60) 
\thinlines
\put(20,70){$(u_1, u_2)$} 
\put(80,70){$T^4$}
\end{picture}}

\put(310,0){\begin{picture}(150,150)
\put(105,120){(c)}
\thinlines
\put(110,10){\line(0,1){100}} 
\put(80,60){\begin{picture}(40,20)
\qbezier(20,0)(20,2)(18,4)
\qbezier(20,0)(20,-2)(18,-4)
\qbezier(-20,0)(-20,-2)(-18,-4)
\qbezier(-20,0)(-20,2)(-18,4)
\qbezier(18,4)(17,6)(14,7)
\qbezier(18,-4)(17,-6)(14,-7)
\qbezier(-18,-4)(-17,-6)(-14,-7)
\qbezier(-18,4)(-17,6)(-14,7)
\qbezier(14,7)(11,8)(8,9)
\qbezier(14,-7)(11,-8)(8,-9)
\qbezier(-14,-7)(-11,-8)(-8,-9)
\qbezier(-14,7)(-11,8)(-8,9)
\qbezier(8,9)(4,10)(0,10)
\qbezier(8,-9)(4,-10)(0,-10)
\qbezier(-8,-9)(-4,-10)(0,-10)
\qbezier(-8,9)(-4,10)(0,10)
\end{picture}} 
\put(80,70){\vector(1,0){0}}

\put(30,60){\begin{picture}(40,20)
\qbezier(20,0)(20,2)(18,4)
\qbezier(20,0)(20,-2)(18,-4)
\qbezier(-20,0)(-20,-2)(-18,-4)
\qbezier(-20,0)(-20,2)(-18,4)
\qbezier(18,4)(17,6)(14,7)
\qbezier(18,-4)(17,-6)(14,-7)
\qbezier(-18,-4)(-17,-6)(-14,-7)
\qbezier(-18,4)(-17,6)(-14,7)
\qbezier(14,7)(11,8)(8,9)
\qbezier(14,-7)(11,-8)(8,-9)
\qbezier(-14,-7)(-11,-8)(-8,-9)
\qbezier(-14,7)(-11,8)(-8,9)
\qbezier(8,9)(4,10)(0,10)
\qbezier(8,-9)(4,-10)(0,-10)
\qbezier(-8,-9)(-4,-10)(0,-10)
\qbezier(-8,9)(-4,10)(0,10)
\end{picture}} 
\put(30,70){\vector(1,0){0}}

\qbezier(40,60)(110,110)(140,60) 
\put(100,85){\vector(1,0){0}}

\qbezier(90,60)(110,70)(140,60) 
\put(120,65){\vector(1,0){0}}

\thicklines
\multiput(20,59)(0,1){3}{\line(1,0){20}} 
\put(18,56){${}_{e_1}$}
\put(34,56){${}_{e_2}$}
\multiput(70,59)(0,1){3}{\line(1,0){20}} 
\put(68,56){${}_{e_3}$}
\put(84,56){${}_{e_4}$}
\multiput(120,59)(0,1){3}{\line(1,0){20}} 
\put(118,56){${}_{e_5}$}
\put(134,56){${}_{\infty}$}
\thinlines
\put(0,60){\line(1,0){150}} 
\put(25,38){$A_1$} 
\put(75,38){$A_2$} 
\put(90,90){$B_1$} 
\put(110,68){$B_2$} 
\put(130,85){$x$} 
\end{picture}}

\put(285,0){\begin{picture}(100,60)
\put(60,90){\line(-1,0){60}} 
\put(10,80){\line(-1,1){10}} 
\put(10,100){\line(-1,-1){10}} 
\put(20,98){$\mu(x)$} 
\put(0,115){Abel-Jacobi map} 
\end{picture}}

\end{picture}
\caption{The genus-$2$ Riemann surface $\Sigma_2$ can be represented as:
(a) $T^2$ cut and glued along two parallel segments at a distance $\alpha$ from each
other; (b) holomorphic curve inside $T^4$; (c) hyperelliptic curve on the $x$-plane,
with branch points at $e_1, e_2, e_3, e_4, e_5, \infty$.
}
\label{fig:curve}
\end{figure}
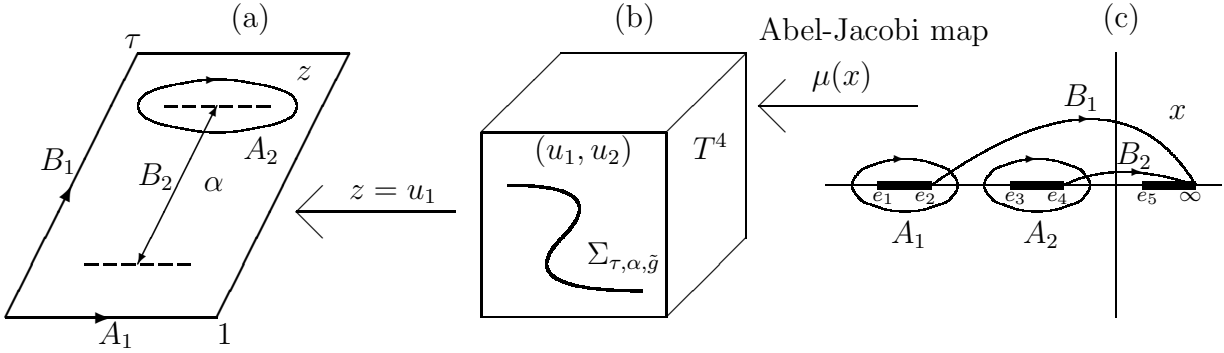

In the denominator, corresponding to the functional determinant, we
split the product over $i\neq j$ into a product over $i<j$ and $i>j$
and relabeled the second product. The measure is now an elliptic
function on $T^2$ with a double zero at $\wl{i}-\wl{j}=0$, and single
poles at $\wl{i}-\wl{j}\pm\tw=0$.  This data determines the measure
uniquely, up to a multiplicative constant.  Recall that the theta
function $\vth_1(z)$ has a simple zero at $z=0$, and that functions of
the form
$$
\prod_i\frac{\vth_1(z-a_i)}{\vth_1(z-b_i)}
$$
are doubly periodic if $\sum_i a_i \equiv \sum_i b_i\, (\mod \Z)$.
Then we can rewrite the above integral as
\bear\label{eq:DVintegral}
\cZ &=& \int d\wl{1}\dots d\wl{\hat N} \prod_{i<j}
  \frac{\vth_1(\wl{i} - \wl{j})^2}
  {\vth_1(\wl{i} - \wl{j} + \tw)
   \vth_1(\wl{i} - \wl{j} - \tw)}
  e^{-\frac{1}{\gST^2}\sum_i W(\wl{i})} \\
  &\equiv&
\int d\wl{1}\dots d\wl{\hat N}
  e^{-\sum_i \mathcal{S}(\wl{i})},
\eear
where
$$
\mathcal{S}(z) \equiv \frac{1}{\gST^2}W(z)
  - \sum_j [2\ln\vth_1(z-\wl{j}) - \ln\vth_1(z-\wl{j} +\tw) - \ln\vth_1(z-\wl{j}-\tw)].
$$
is the effective action of a probe eigenvalue placed at $z$.  As usual,
the resolvent $R(z)$ is defined to be the force on a probe eigenvalue due to
all the other eigenvalues and is given by
$$
R(z)=\gST^2 \sum_j\frac{\vth_1^\prime(z-\wl{j})}{\vth_1(z-\wl{j})}.
$$
In the large $\hat N$ limit, the eigenvalues condense into cuts on
the $z$-plane and the saddle point equation for the eigenvalues is
expressed in terms of the resolvent and $S \equiv g_{st}^2 \hat N$ as
\be\label{saddle}
W^{\prime}(z) - S\left(2R(z) - R(z+\tw) - R(z-\tw)\right) = 0,
\ee
whenever $z$ is on a cut.
Following \cite{Hollowood:2003gr}, if we can express $W(z)$ as
\be\label{eq:WUrel}
W(z) = U(z+\frac{\tw}{2}) - U(z-\frac{\tw}{2}),
\ee
for some possibly quasiperiodic function $U(z)$ on $\Sigma_\tau$,
we can re-express (\ref{saddle}) as
\be\label{branch-cut}
J(z+\frac{\tw}{2} \pm i\epsilon)
  = J(z - \frac{\tw}{2} \mp i\epsilon),
\ee
where $J(z)$ is the auxiliary function
\be\label{eq:mmJ}
J(z) = U^\prime(z) + S[R(z+\frac{\tw}{2}) - R(z-\frac{\tw}{2})].
\ee
We note that only $U^\prime(z)$, not
$U(z)$, need be elliptic.  This reformulation of the saddle
point equation states that $J(z)$, which is already doubly periodic, is
discontinuous at the cuts.  In fact
the saddle point equation implies that $J(z)$ is a function on a
genus $2$ Riemann surface $\Sigma_2$ obtained from the
(dual) $T^2$ of complex structure $\tau$ by cutting it along two
segments and gluing the cuts to each other.  As depicted in figure
(\ref{fig:curve}), the top side of the upper cut is glued to the 
bottom side of the lower cut and vice versa, yielding two new cycles
$A_2$ and $B_2$.  

\par Given the function $J(z)$ on $\Sigma_2$ one can complete the DV
approach by writing down the gaugino superpotential as follows.  The
single cut solution of the matrix model corresponds in the physical
gauge theory to a classically unbroken $SU(N)$ vacuum which then
quantum mechanically confines, generating a mass gap and  yielding 
a gaugino condensate $S$.  In the
matrix model, $S$ is given by
\be
S = - \frac{1}{2\pi} \int_{A_2} J(z) dz.
\ee
The matrix model free energy in the planar, large $\hat N$ limit 
is then given by
\be
\frac{\partial {\cal F}_0}{\partial S} = - i \int_{B_2} J(z) dz,
\ee
From these expressions, one finally arrives at the DV gaugino superpotential
\be\label{eq:DVsuppot}
W_{\mathrm{eff}}(S) = N \frac{\partial {\cal F_0}}{\partial S} - 2\pi i \rho S,
\ee
where $\rho$ is the effective 3+1D $SU(N)$ bare (complex) gauge
coupling.
To obtain a quantitative check of the DV approach for the compactified 6D
theory, one could calculate the value of the superpotential \eqref{eq:DVsuppot}
at its minimum and compare to the Seiberg-Witten theory or integrable 
systems approaches.  In order to do this, it is important to have an
explicit construction of the genus 2 matrix model spectral curve and
functions on it.

\vskip 0.5cm
\vskip 0.5cm

\par Any genus $g$ Riemann surface $\Sigma_g$ can be embedded in its
Jacobian $T^{2g}$ by the Abel-Jacobi map $\mu:\Sigma_g \rightarrow
T^{2g}$ (see appendix \ref{sec:ThetaApx} for some details).  In our case
the Jacobian is a $T^4$, whose complex structure is determined by the
period matrix of $\Sigma_2$, which we can always choose to be
\be\label{eq:SCperiods} 
\left(\begin{array}{cccc} 1 & 0 & \Omega_{11} & \Omega_{12}
\\ 0 & 1 & \Omega_{12} & \Omega_{22} \\
\end{array}\right).
\ee 
The columns represent the four 1-cycles $A_1, B_1, A_2, B_2$
of $\Sigma_2$ in that order, and the rows represent the two holomorphic
1-forms $\omega_1, \omega_2$.  Apriori, the $\Omega_{ij}$ are arbitrary,
but we will determine them soon.  We choose complex coordinates $u =
(u_1, u_2)$ on $T^4$.  For the special case of genus 2, the image of
$\Sigma_2$ under the Abel-Jacobi map $\mu$ sits inside the $T^4$ as the
zero locus of a particular theta function which we call $\vth_0$ \cite{Tata}:
\be
\mu(\Sigma_2) = \bigg\{ \vth_0(u|\Omega) \equiv \thxx{1/2}{1/2}{0}{1/2}(u|\Omega) = 0 \bigg\}.
\ee
$\vth_0(u)$ has the nice property that it is odd under the $\Z_2$
action $u \rightarrow -u$, which immediately implies that $\mu(\Sigma_2)$
enjoys the same $\Z_2$ involution symmetry.

\vskip 0.5cm

\par Important constraints on the periods of the $B$-cycles of
$\Sigma_2$, namely the $2 \times 2$ matrix $\Omega$, can be obtained by
relating the $T^4$ embedding of $\Sigma_2$ to its presentation in
figure (\ref{fig:curve}a) as a torus on the z-plane with two cuts
glued together.  This presentation implies that there exists a
holomorphic, but quasiperiodic function $z$ on $\Sigma_2$ with no poles.  Such a
function can be constructed as a quasiperiodic function on $T^4$ 
restricted to $\mu(\Sigma)$.  Let $\pi_i$ , $i=1\dots4$, be the columns
of the period matrix \eqref{eq:SCperiods}.  Then $z(u)$ obeys the periodicities
\bear
z({u} + {\pi}_1) &=& z({u}) + 1 \\
z({u} + {\pi}_2) &=& z({u}) \\
z({u} + {\pi}_3) &=& z({u})+ \tau \\ 
z({u} + {\pi}_4) &=& z({u}) + \alpha.  
\eear
The only functions on $T^4$ that have no poles, even when restricted
to $\mu(\Sigma_2)$, are $u_1$ and $u_2$.  Without loss of generality we
choose the function $z$ to be $u_1$.  Identifying $z$ with $u_1$
implies that we must have $\Omega_{11} = \tau$, and $\Omega_{12} = \alpha$
in order to satisfy the periodicities above.  Hence our matrix model
spectral curve has only one undetermined complex structure modulus $\Omega_{22}$.
In the matrix model solution this is related to the size of the cut, while
in the physical gauge theory, it is related to the gaugino condensate $S$.
Only after minimizing the DV superpotential \eqref{eq:DVsuppot} will
we be able to determine
$\Omega_{22}$.  Indeed we shall find that $\Omega_{22} = \frac{\rho}{N}$.

\par
Having pinned down somewhat the matrix model spectral curve, we now turn to the
description of the function $J(z)$.  It may help to retrace the steps leading
to the definition \eqref{eq:mmJ} of $J(z)$ with a specific \SUSY{1} deformation
$W$ in mind as an example.  For convenience, we choose the deformation
\be
W(z) = \zeta(z+\alpha) - \zeta(z-\alpha).
\ee  
This is an elliptic function on
the compactification torus $\Sigma_\tau$, the space on which the eigenvalues of 
the matrix model live.  The eigenvalues like to sit near the critical points
of $W$ given by 
\be
\partial_z W = - \wp(z+\alpha) + \wp(z-\alpha) = 0.
\ee
This equation is satisfied for $z=0$ because $\wp(z)$ is an even function so 
we will look for a single cut solution where the eigenvalues condense around $z=0$.
Next we need to find a function $U(z)$ satisfying \eqref{eq:WUrel}.  Given
our choice of $W$, \eqref{eq:WUrel} is trivially satisfied by $U(z) = \zeta(z)$.
$U(z)$ is only a quasiperiodic function on $\Sigma_\tau$, but its derivative
$U^\prime(z) = - \wp(z)$ is nevertheless fully periodic.  So finally
$J(z)$ in \eqref{eq:mmJ} is given by
$$
J(z) = - \wp(z) + S[R(z+\frac{\tw}{2}) - R(z-\frac{\tw}{2})],
$$
and via the saddle point equation \eqref{branch-cut} is a function on the
genus 2 spectral curve $\Sigma_2$.  The function is determined by its
pole structure, specified by $U^\prime(z)$ and its symmetries.  We note
that $J(z)$ has a single pole on $\Sigma_2$ of order 2.

\par In order to calculate the DV superpotential \eqref{eq:DVsuppot}, we must
fix the one form $J(z)dz$ and calculate its periods on $\Sigma_2$. In doing 
this it can be
helpful to use a hyperelliptic representation of $\Sigma_2$.
Any genus-$2$ Riemann surface can be represented as a hyperelliptic curve,
i.e. a double cover of the complex plane (see figure \ref{fig:curve}):
\be\label{yyxxxxx}
y^2 = \prod_{i=1}^5 (x - e_i).
\ee
A basis for the holomorphic 1-forms on $\Sigma_2$ is given by
$$
\omega_1\equiv\frac{dx}{y},\qquad \omega_2\equiv\frac{x dx}{y}.
$$
At $x=\infty$ a good coordinate is $w\equiv x^{-\frac{1}{2}}$ so that
$x=1/w^2$ and $y\sim 1/w^5$ and the above forms are nonsingular at $w=0$.

To proceed, we need a meromorphic 1-form with a single singularity of degree $2$.
The form 
$$
J = \frac{x^2 dx}{y}
$$
has a double pole at $x=y=\infty$ and is the 1-form we are looking for.

We can also express $J$ in terms of the
embedding $i:\Sigma_2\hookrightarrow T^4.$
To do this we identify
$$
\omega_1 = \frac{dx}{y} = i^* du_1,\qquad
\omega_2 = \frac{x dx}{y} = i^* du_2.
$$
On $\Sigma_2$ we have
$$
\px{1}\vth_0 du_1 + \px{2}\vth_0 du_2 = 0\Longrightarrow
\left.\frac{du_2}{du_1}\right|_{\Sigma_2} = 
\left.-\frac{\px{1}\vth_0}{\px{2}\vth_0}\right|_{\Sigma_2},
$$
We can therefore set
$$
x = \left.\frac{\omega_2}{\omega_1}\right|_{\Sigma_2}
 = \left. -\frac{\px{1}\vth_0}{\px{2}\vth_0}\right|_{\Sigma_2}
$$
and
$$
J =\left. -\frac{\px{1}\vth_0}{\px{2}\vth_0} du_1\right|_{\Sigma_2} .
$$

\par It can be daunting to evaluate the period integrals of $J(z)$
directly, but as 
explained in \cite{Hollowood:2003cp}, one can nevertheless find the moduli of
the matrix model spectral curve at a critical point of the DV
superpotential indirectly. 
The crucial insight is to realize that the 1-form
\be\label{eq:actholo}
\beta = - \frac{1}{2\pi} \frac{\partial}{\partial S} J(z) dz
\ee
is actually a holomorphic 1-form.  This is because the only singularity
in $J(z)$, coming solely from $U^\prime(z)$, is manifestly independent of
the modulus $S$ associated to the cut.  We know the integral of this one
form over the $A_1$ cycle in figure (\ref{fig:curve}) is $0$, since
the integral 
$\int_{A_1}J(z)dz$ is identically $0$ independent of $S$.  Therefore $\beta$, 
in the $T^4$ language,  cannot
be $du_1$ and hence must be $du_2$.  Now the extremum condition for the 
gaugino superpotential \eqref{eq:DVsuppot} is
\be\label{eq:critical1}
N \frac{\partial^2 {\cal F}_0}{\partial S^2} = 2 \pi i \rho.
\ee
Using the relation \eqref{eq:actholo} with $\beta = du_2$, we can 
rewrite
\be\label{eq:critical2}
\frac{\partial^2 {\cal F}_0}{\partial S^2} = 2 \pi i \int_{B_2}du_2 = 
2 \pi i\, \Omega_{22}.
\ee
Thus, as promised, at the critical point of the gaugino
superpotential, the one undetermined modulus $\Omega_{22}$ of the
spectral curve is fixed by \eqref{eq:critical1} and
\eqref{eq:critical2} to be $\frac{\rho}{N}$.  Hence the period matrix
\eqref{eq:SCperiods} for the spectral curve at the critical point becomes
\be
\left(\begin{array}{cccc} 
1 & 0 & \tau & \alpha           \\
0 & 1 & \alpha & \frac{\rho}{N} \\
\end{array}\right).
\ee
 
This is very similar to the period matrix \eqref{T4periods} of the
physical $T^4$ in which the Seiberg-Witten curves of the undeformed
\SUSY{2} theory live.  Indeed the moral so far of the six-dimensional
DV story is that the Jacobian of the genus 2 matrix model spectral
curve associated to a one-cut solution, is closely related to the
physical $T^4$ in which the Seiberg-Witten curves of the undeformed
\SUSY{2} theory live.  We will investigate this relation in more
detail in the next section, but for now we show, again following
\cite{Hollowood:2003cp}, that the evaluation of the DV superpotential
at its critical point can be reduced to the evaluation of a residue.

\par
Assembling previous results, we have
\bear
W_{eff} &=& N \frac{\partial {\cal F}_0}{\partial S} - 2\pi i \rho S    \\
        &=& iN \bigg(-\int_{B_2} J(z) dz + \frac{\rho}{N} \int_{A_2} J(z)dz \bigg) \\
        &=& iN \bigg(-\int_{A_2}du_2 \int_{B_2} J(z) dz + 
                     \int_{B_2}du_2 \int_{A_2} J(z)dz \bigg) \\
        &=& \mathrm{Res}_{z \rightarrow P}\bigg(U(z)u_2\, dz\bigg).
\eear
In the last line we used a Riemann bilinear relation assuming $U(z)$
has no simple poles.  Here $P$ is the location of the higher order
pole of $U(z)$.  Fortunately, no daunting integrals need be calculated
to extract physical information from the matrix model.

\section{The DV matrix integrals for $c_1\neq 0$}\label{sec:DVSP}

We will now study the \SUSY{1} compactifications described in subsection
(\ref{subsec:chern}).
We take the chiral field multiplet $\Phi_{+}$ to live in a line
bundle $\LBd$ on $T^2$ with a nonzero first Chern class $c_1(\LBd) = k.$
The other chiral multiplet $\Phi_{-}$ takes values in $\LBd^{-1}.$
For simplicity, we set $k=1.$
To specify the field theoretic action we must pick a connection
$$
\rB = \rB_z dz + \rB_\bz d\bz
$$
for $\LBd$ on $T^2.$

The Matrix Model integral (\ref{DVmm}) becomes
$$
\int [D\cP_{+}][D\cP_{-}][D\cA_\bz]
e^{-\int d^2z\, \tr \{\Phi_{-} D_\bz\Phi_{+} -i\cP_{+}[\cA_\bz,\cP_{-}]\}}
=\int [D\cA_\bz]\frac{1}{\det (D_\bz + i [\cA_\bz,\cdot])}
$$
Here $D_\bz =\px{\bz} + i \rB_\bz.$

To calculate the determinant, we represent $T^2$ explicitly with
variables $0\le x_1, x_2\le 1$ and
$$
z = x_1 + \tau x_2,\qquad \bz = x_1 + \btau x_2,\qquad
D_\bz = \frac{1}{2i\tau_2}(\tau D_1-D_2),
$$
where
$$
D_1\equiv \px{1} + i \rB_1,\qquad D_2 \equiv \px{2} + i \rB_2,\qquad
\tau \equiv \tau_1 + i\tau_2.
$$
We expand
\be\label{hPhiDec}
\Phi_{+}(x_1, x_2) = \sum_{q=0}^{k-1}
\sum_{n=-\infty}^\infty e^{-2\pi i (n k + q) x_1}\hPhi^{(q)}_{+}(x_2 - n)
\stackrel{k=1}{\longrightarrow}
\sum_{n=-\infty}^\infty e^{-2\pi i n x_1}\hPhi_{+}(x_2-n)
\ee
This expansion obeys 
the boundary conditions for the line bundle $\LBd$
$$
\Phi_{+}(x_1 + 1, x_2) = \Phi_{+}(x_1, x_2),\qquad
\Phi_{+}(x_1, x_2 + 1) = e^{-2\pi i k x_1}\Phi_{+}(x_1, x_2).
$$
The whole information present in the two-dimensional field $\Phi_{+}$ is 
thus encoded in the $k$ one-dimensional fields $\hPhi^{(q)}$ ($q=0\dots k-1$).
We take the gauge field to be
$$
\rB = 2\pi k x_2 dx_1 \stackrel{k=1}{\longrightarrow} 2\pi x_2 dx_1
$$
We will intially restrict to $k=1$ completely; the generalization to
positive $k$ is straightforward.
It is convenient to define a Heisenberg algebra
$$
\ha\equiv \frac{1}{\sqrt{4\pi\tau_2}}(i\px{2} + 2\pi \tau x_2),\qquad
\hadg = \frac{1}{\sqrt{4\pi\tau_2}}(i\px{2} + 2\pi \btau x_2),\qquad
[\ha,\hadg] = 1. 
$$
Then
$$
D_\bz\Phi_{+}(x_1, x_2) =
\sqrt{\frac{\pi}{\tau_2}}
\sum_{n=-\infty}^\infty e^{2\pi i n x_1}[\ha\hPhi_{+}](x_2-n)
$$
Thus, formally,
\be\label{nonsense}
\det (D_\bz + i[\cA_\bz,\cdot]) = \prod_{i\neq j}
\det \left[(\wl{i} -\wl{j})\hI + \sqrt{\frac{\pi}{\tau_2}}\ha\right]
\ee
Here $\hI$ is the identity operator on the representation of the Heisenberg algebra.

Note that since the periodicities of $A$ are
\be
A_\bz \cong A_\bz + \frac{\pi}{\tau_2}
      \cong A_\bz + \frac{\pi\tau}{\tau_2},
\ee
the eigenvalues of $A\bz$ take values in a torus with similar periodicities.
Then, including the Jacobian, the Matrix model integral (\ref{DVmm}) becomes
\be\label{Zchn}
\cZ=\int d\lambda_1\dots d\lambda_{\hat N}\prod_{i\neq j}
  \frac{\prod_{m,n}(\wl{i}-\wl{j} + m\frac{\pi}{\tau_2}
                                  + n\frac{\pi\tau}{\tau_2})} 
       {\det \left[(\wl{i} -\wl{j})\hI + \sqrt{\frac{\pi}{\tau_2}}\ha\right]}
\ee
As it stands, the denominator of (\ref{Zchn}) does not make sense.
The term containing $\ha$ can seemingly be dropped since it does not affect
the determinant (being an upper triangular matrix in the harmonic operator
representation).

We would like to propose the following regularization of (\ref{Zchn}).
Define the basis of coherent states:
$$
\ket{\zeta} = e^{\hadg \zeta -\frac{1}{2}|\zeta|^2}\ket{0},
$$
where $\ket{0}$ is the ground state of the Heisenberg algebra ($\ha\ket{0} = 0$).
Then, using
$$
\tr \mathcal{O} = \frac{1}{\pi}\int d^2\zeta \bra{\zeta}\mathcal{O}\ket{\zeta},
$$
we get
\bear
\log\det \left[(\wl{i} -\wl{j})\hI +\sqrt{\frac{\pi}{\tau_2}}\ha\right]
&=&
\tr \log \left[(\wl{i} -\wl{j})\hI +\sqrt{\frac{\pi}{\tau_2}}\ha\right]
\nn\\
&=&
\frac{1}{\pi}\int d^2\zeta \log \left(\wl{i} -\wl{j} 
    + \sqrt{\frac{\pi}{\tau_2}}\zeta\right)
\nn
\eear
We redefine $\zeta \rightarrow \sqrt{\frac{\pi}{\tau_2}}$.
At this point we would like to replace the $d^2\zeta$ integration
over the whole $\C$ plane by a sum over integer pairs $(n,m)$ and an integral
over a fundamental domain that is a $T^2$ with sides
$\frac{\pi}{\tau_2}$ and $\frac{\pi\tau}{\tau_2}$.
$$
\zeta = n\frac{\pi}{\tau_2} + m\frac{\pi\tau}{\tau_2} + \eta,\qquad
\mathrm{ with }\, \eta\in T^2.
$$
and for any expression $F(\zeta)$ we replace
$$
\int_\C F(\zeta) d^2\zeta \longrightarrow 
\sum_{n,m}\int_{T^2} d^2\eta F(\eta + m\frac{\pi}{\tau_2} +
                                      n\frac{\pi\tau}{\tau_2}).
$$

We obtain
$$
\int d^2\zeta \log \left(\wl{i} -\wl{j} +\zeta\right)
=\sum_{n,m} \int_{T^2} d^2\eta
\log \left(\wl{i} -\wl{j} + m\frac{\pi}{\tau_2}
                          + n\frac{\pi\tau}{\tau_2} + \eta\right)
$$
The motivation behind this regularization becomes apparent if recall
that large values of $\zeta$ can be interpreted classically.
For example if $\zeta \approx m'\frac{pi}{\tau_2}
 + n'\frac{\pi\tau}{\tau_2}$ then the corresponding wave-function
$\hPhi^{(q)}(x_2)$ is localized near $x_2\approx m'$.
The dominant term in (\ref{hPhiDec}) will then have $n\approx n'.$
For large values of $(m',n')$ the wave function will therefore behave
like $\exp (i m' x_1 + i n' x_2).$
Therefore, it makes sense to rewrite the denominator of
(\ref{Zchn}) so that terms
with $\zeta$'s near $m\frac{\pi}{\tau_2}+n\frac{\pi\tau}{\tau_2}$
should combine with the term
$(\wl{i}-\wl{j} + m\frac{\pi}{\tau_2} + n\frac{\pi\tau}{\tau_2})$ in
such a way that the total product will be finite.

Using this, and noting that
$\mathrm{vol}(T^2)=\int_{T^2}d^2\eta=\frac{\pi^2}{\tau_2}$, we can find

\bear\nonumber
  & &\prod_{i\ne j} \frac{
           \prod_{m,n}(\wl{i}-\wl{j} + m\frac{\pi}{\tau_2}
                                     + n\frac{\pi\tau}{\tau_2}) }
     {\det \left[(\wl{i} -\wl{j})\hI + \sqrt{\frac{\pi}{\tau_2}}\ha\right]}
  = \\
& &\quad\exp\left[\frac{\tau_2}{\pi^2} \int_{T^2}d^2\eta \sum_{i<j}
  \log \frac{\vth_1^2(\lambda_i - \lambda_j)}
    {\vth_1(\lambda_i-\lambda_j -\eta)\vth_1(\lambda_i-\lambda_j +\eta)}
\right]
\eear

\noindent Inserting the $\tw$-twists from our previous case, and
expressing the K\"ahler modulus of the eigenvalue torus as
$\beta=\pi^2/\tau_2$, we obtain
\be\label{Zchfinal}
  \cZ=\int d\lambda_1\dots d\lambda_{\hat N}
  \prod_{i<j} \exp\left[\frac{1}{\beta} \int_{T^2}d^2\eta
    \log \frac{\vth_1^2(\lambda_i - \lambda_j)}
    {\vth_1(\lambda_i-\lambda_j -\alpha -\eta)
     \vth_1(\lambda_i-\lambda_j +\alpha +\eta)}
  \right].
\ee

\vskip 0.5cm
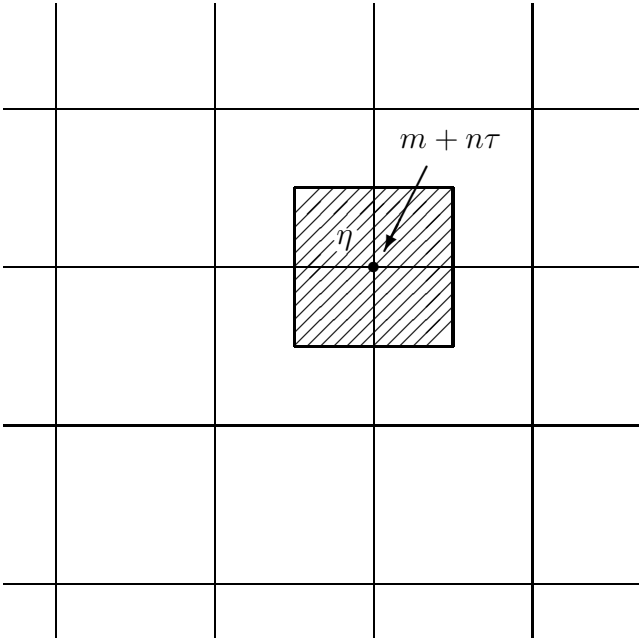
\begin{figure}[t]
\begin{picture}(400,250)
\thinlines
\multiput(0,20)(0,60){4}{\line(1,0){240}}
\multiput(20,0)(60,0){4}{\line(0,1){240}}
\thicklines
\put(150,185){$m + n\tau$}
\put(160,178){\vector(-1,-2){16}}
\put(140,140){\circle*{4}}
\put(126,149){$\eta$}
\multiput(110,110)(0,60){2}{\line(1,0){60}}
\multiput(110,110)(60,0){2}{\line(0,1){60}}


\thinlines

\put(110,115){\line(1,1){55}}
\put(110,120){\line(1,1){50}}
\put(110,125){\line(1,1){45}}

\put(110,130){\line(1,1){16}} 
\put(150,170){\line(-1,-1){16}} 

\put(110,135){\line(1,1){15}} 
\put(145,170){\line(-1,-1){15}} 

\put(110,140){\line(1,1){30}}
\put(110,145){\line(1,1){25}}
\put(110,150){\line(1,1){20}}
\put(110,155){\line(1,1){15}}
\put(110,160){\line(1,1){10}}
\put(110,165){\line(1,1){5}}

\newcounter{Xco}
\newcounter{Yco}
\newcounter{TLen}
\setcounter{Xco}{110}
\setcounter{Yco}{110}
\setcounter{TLen}{60}
\loop
\put(\value{Xco},\value{Yco}){\line(1,1){\value{TLen}}}
\addtocounter{Xco}{5}
\addtocounter{TLen}{-5}
\ifnum \value{TLen}>0
\repeat

\end{picture}
\caption{The ill-defined expression (\ref{Zchn}) is regularized in
(\ref{Zchfinal})
by splitting the denominator into little pieces, each being an integral
over a $T^2$ centered around the lattice point $m+n\tau.$
}
\end{figure}
\vskip 0.5cm

Note that this matrix integral is similar to the usual one
\eqref{eq:DVintegral} except for the fact that an \SUSY{1} term is not
added by hand.  Instead the effects of the SUSY breaking nontrivial
R-symmetry Chern class manifests itself in the matrix integral as a
smeared background charge in which the eigenvalues float.  It would be
interesting to analyze this integral further, but for now we turn to an
analysis of these compactified six-dimensional theories from the
Seiberg-Witten and integrable systems viewpoint.

\section{The Integrable Systems Approach}\label{sec:intsys}

A useful point of view in Seiberg-Witten theory is to view the total
space $\cal M$ of Jacobians of the Seiberg-Witten curve as the phase
space of an algebraic integrable system \cite{DonagiWitten:1995}.
More specifically, $\cal M$ is a fibration of abelian varieties
(complex projective tori) with base space $\Msw$, the moduli space of
Seiberg-Witten curves.  If $\Sigma$ is a Seiberg-Witten curve with
some particular moduli $p \in \Msw$, the fiber $\pi^{-1}(p)$ of the
projection $\pi : \cal M \rightarrow \Msw$ is simply the Jacobian
\Jac{\Sigma} of $\Sigma$.  If $\Sigma$ is a curve of genus $g$ then
\Jac{\Sigma} is a complex $g$ dimensional abelian variety.  As is well
known, the complex structure of \Jac{\Sigma} controls the holomorphic
couplings of the $g$ $U(1)$ photons present at generic points in the
moduli space of vacua of the $4D$ theory.

\par The utility of this viewpoint is made manifest upon compactifying
the 4D theory on a circle $S^1$ of radius $R$ to obtain a 3D theory
with $8$ supercharges.  While the moduli space of vacua of the 4D
theory is $\Msw$, the moduli space of vacua of the 3D theory is
enlarged to $\cal M$ \cite{SeibergWitten:1996}.  The extra complex
scalars in 3D come from dimensionally reducing the 4D vector, and are
comprised of Wilson loops and dual photons.  The holomorphic couplings
of the 4D $U(1)$ photons reduce to kinetic terms for these 3D scalars,
and hence the Jacobian of the Seiberg-Witten curve becomes part of the
moduli space of the 3D theory.  This moduli space of vacua is
hyperkahler, but its complex structure can be chosen to be independent
of $R$ \cite{SeibergWitten:1996}.

\par After softly breaking the $4D$ \SUSY{2} theory by an \SUSY{1}
superpotential $W$, the compactified $3D$ theory also develops a
superpotential $\cal W$.  The $3D$ superpotential $\cal W$ is a
meromorphic function on $\cal M$, which as noted above is the phase
space of an integrable system.  It turns out that $\cal W$ is
generically some combination of the $\frac{1}{2} \mathrm{Dim}_{\mathbb
C} \cal M$ Poisson commuting Hamiltonians associated with the
integrable system on $\cal M$.  The relationship between the \SUSY{1}
deformation $W$ and the $3D$ superpotential $\cal W$ can be made
explicit for supersymmetric gauge theories when the associated
integrable system has a Lax matrix formulation \cite{BBDW:2003}.

\par After such a relationship has been identified, one can ask for
the supersymmetric vacua of the $3D$ theory, given by extrema of $\cal
W$ on $\cal M$.  In the integrable systems language, these extrema are
nothing more than fixed points of the Hamiltonian flow on $\cal M$
generated by $\cal W$.  In other words there is a correspondence
between equilibrium configurations of the integrable system with
respect to the Hamiltonian $\cal W$ and supersymmetric vacua surviving
the \SUSY{1} deformation $W$ \cite{Hollowood:2003cp}.

\par Furthermore, as explained in \cite{Hollowood:2003cp},
the integrable systems approach also sheds light on the alternate
Dijkgraaf-Vafa approach to calculating the vacuum structure of the
\SUSY{1} deformed theory.  Briefly, at an equilibrium point $p \in \cal
M$ of the Hamiltonian $\cal W$, the Seiberg-Witten curve associated to
that point can be shown to degenerate to a curve of lower genus.  At
massive vacua, the curve maximally degenerates.  In the Dijkgraaf-Vafa
approach, minimization of the gluino superpotential amounts to a
condition on the matrix model spectral curve that relates its moduli
to that of the maximally degenerated Seiberg-Witten curve.  Again see
\cite{Hollowood:2003cp} and references therein for more details.

\par These beautiful relationships between massive vacua of \SUSY{1}
supersymmetric gauge theories, equilibrium points of integrable
systems, maximally degenerated Seiberg-Witten curves, and matrix model
spectral curves are expected to also hold for the $6D$ twisted,
compactified gauge theories considered in this paper.  This is
plausible because the $4D$ and $5D$ cases in which evidence for such
relationships have been accumulated \cite{Dorey:2002, Dorey:1999,
BBDW:2003} are all continuously connected to the $6D$ case via
degenerations. In addition, integrable systems for other $5D$ and $6D$ theories
have been found in \cite{Nekrasov:1996cz,Gorsky:1997mw}.
A critical step towards obtaining these results in
$6D$ is an understanding of the underlying integrable system governing
the theory, a task to which we now turn.

\par In order to understand the integrable system, we first review and
reformulate what is known about the Seiberg-Witten curves of the 6D
theory, which were found in \cite{Ganor:2000un, Cheung:1998wj} as
spectral curves of noncommutative instantons on $T^4$ via string
duality arguments.  For $SU(N)$ gauge theory the Seiberg-Witten curves
are divisors, or zero loci of holomorphic sections, of a line bundle
$\LBd$ over $T^4$.  If we choose real coordinates $x_1,\dots,x_4$ on
$T^4$ such that $0\le x_i\le 1$ and periodic boundary conditions given
by $x_i\sim x_i+1$, then the Chern class $\psi$ of $\LBd$ is given by

\be\label{ChernClass}
\psi = dx_1\wedge dx_3 + N \, dx_2\wedge dx_4.
\ee

Choosing holomorphic coordinates $u_1$ and $u_2$, the period matrix of
$T^4$ (integrals of $du_1, du_2$ over the 4 1-cycles of $T^4$) is
expressed in terms of the compactification data as

\be\label{T4periods}
\left(\begin{array}{cccc}
N & 0 & \tau & \alpha \\
0 & 1 & \alpha & \rho \\
\end{array}\right), \qquad 
\ee

\noindent where $\tau$ is the complex structure of the $T^2$
compactification, $\alpha$ is the twist parameter, and $\rho$ is the
$4D$ effective complex gauge coupling
\cite{Ganor:2000un, Cheung:1998wj}.
$\LBd$ has precisely $N$ sections which we call $\vth_i$, for $i =
0,\dots,N-1$.   
In terms of the theta functions defined in appendix
\ref{sec:ThetaApx}, we can choose 
$\vth_i$ to be
$$
\vth_i \equiv \thxx{i/N}{0}{0}{0}(u|\Omega).
$$
With this data, the Seiberg-Witten curves are written explicitly as zero loci
of sections of $\LBd$ given by the equation
\be\label{eq:SWcurve}
\sum_{i=1}^{N-1} a_i\vth_i = 0.
\ee

\noindent Each of these Seiberg-Witten curves have genus $N+1$,
as can be derived from
the adjunction formula, using the fact that divisors of $\LBd$
are Poincar\'e dual to the Chern class $\psi$.
 The moduli space of Seiberg-Witten curves
$\Msw$ is given by the space of complex coefficients
$a_0,\dots,a_{N-1}$ modulo overall rescalings and hence is $\P^{N-1}$.

\par The complex $2N-2$ dimensional phase space $\cal M$ of the
underlying integrable system should be the total space of a family of
complex $N-1$ dimensional abelian varieties fibered over $\Msw =
\P^{N-1}$.  If $p \in \P^{N-1}$, and $\Sigma$ is the Seiberg-Witten
curve associated to $p$, then the $N-1$ dimensional abelian
variety sitting above $p$ in $\cal M$ can be obtained from the part of
\Jac{\Sigma} that does not come from $T^4$.  More precisely, the two
holomorphic one forms $du_1$, $du_2$ on $T^4$ pullback to holomorphic
one forms $\omega_1$, $\omega_2$ on $\Sigma$ via its embedding map
into $T^4$.  These one forms in turn span a complex 2-dimensional
subspace of the $N+1$ dimensional abelian variety \Jac{\Sigma}.
Modding out by this subspace yields the $N-1$ dimensional variety we
seek.

\par One can think of coordinates on $\P^{N-1}$ and coordinates on the
$N-1$ dimensional abelian varieties as the action and angle variables
respectively of some system of particles.  However, in order to gain
intuition for the above rather abstractly presented integrable system,
it would be useful to actually have a particle-like interpretation of
its degrees of freedom, and an explicit formula for its $N-1$ Poisson
commuting Hamiltonians.  This would be a likely first step in finding
equilibrium configurations and connecting to the DV approach. The
technique we use to obtain this particle-like interpretation is the
separation of variables \cite{Sklyanin:1995}.  The geometry behind
this approach was explained in \cite{GNR:1999} which we now briefly
review.

\par Given a surface $S$ which we can take generally to be a $K3$
manifold, one can define an integrable system $\cal M$ whose Poisson
commuting action variables parameterize the moduli space of holomorphic
curves $\Sigma$ dual to a fixed cohomology class $\psi$ and hence of
fixed genus $g$.  The angle variables will parameterize the moduli
space of line bundles $\LBd$ of degree $g$ on $\Sigma$.  This moduli
space for a given $\Sigma$ is just \Jac{\Sigma}.  A key observation is
that generically one can map the total space $\cal M$ of this
integrable system to \Symm{^gS} as follows.  A generic section of
$\LBd$ has $g$ zeros.  Simply map the point $(\Sigma, \LBd)$ of $\cal
M$ to the point in \Symm{^gS} represented by these $g$ zeroes.  The
inverse map can also be constructed.  Given $g$ points in $S$, there
is generically a unique curve $\Sigma$ dual to the cohomology class $\psi$
passing through those points, with a line bundle $\LBd$ having those
$g$ points as a divisor.  Furthermore both these maps are
symplectomorphisms, where the symplectic structure on \Symm{^gK3} is
the natural one induced by that on $K3$.  Physically, the curve
$\Sigma$ along with line bundle $\LBd$ can be thought of as a bound
state of $D2$ branes in $S$, and the map to $g$ points, or $D0$ branes
is a T-duality map, or at the level of sheaves a Fourier-Mukai
transform.  The system of $g$ points on $S$ constitute the
separated variables of the integrable system $\cal M$.

\par Returning to our case, we have $S = T^4$ but we do not have every
curve Poincar\'e dual to $\psi$ in \eqref{ChernClass} at our disposal.
We have only those curves that are divisors of the fixed line bundle
$\LBd$ on $T^4$.  Generically a translate in $T^4$ of such a divisor
is no longer a divisor of $\LBd$.  Motivated by the separation of
variables technique, we would like to map our integrable system $\cal
M$ to a system of points on $T^4$ \footnote{This point was
independently noted in \cite{Hollowood:2003cv}.}.  Physically, we
require a $2N-2$ dimensional phase space, indicating that our
integrable system is really a collection of $N-1$ identical particles
moving on $T^4$.  We now check that one can construct the inverse map:
given $N-1$ points there is a unique divisor $\Sigma$ of $\LBd$ going
through those $N-1$ points.

\par To see this, it is useful to consider the canonical embedding
$\iota : T^4 \rightarrow \P^{N-1}$ using the space of sections of
$\LBd$.  Given a point $x \in T^4$, $\iota(x)$ is a point in
$\P^{N-1}$ specified by the homogeneous coordinates $[\vth_0(x),
\dots, \vth_{N-1}(x)]$.  As long as there is no $x \in T^4$ for
which $\vth_i(x) = 0\ \forall\ i$ this map will be well defined.
This consistency condition will be met for $N$ sufficiently large.
The utility of the canonical embedding $\iota$ is that the space of
Seiberg-Witten curves can now be viewed as the space of codimension 1
hyperplanes in $\P^{N-1}$.  Any such hyperplane slices the embedding
of $T^4$ in $P^{N-1}$ in a Seiberg-Witten curve $\Sigma$.
Furthermore, any $N-1$ points in general position in $P^{N-1}$
trivially yield a unique hyperplane through those points.  If in
addition these points are on the image of $T^4$, this hyperplane
carves out the unique Seiberg-Witten curve in $T^4$ going through
those points.  As these $N-1$ points move around on $\Sigma$, they
also parameterize the $N-1$ angle variables of the system.

\par We are now in a position to explicitly write down the $N-1$
Poisson commuting Hamiltonians of our integrable system as a function
of the separated variables consisting of the $N-1$ points $x_1, \dots,
x_{N-1}$ in $T^4$.  We know from above that given generic $x_1, \dots,
x_{N-1}$, we can determine $a_0, \dots a_{N-1}$ uniquely by the
condition that $x_1, \dots, x_{N-1}$ lie on the same curve, specified
by equation \eqref{eq:SWcurve}.  Furthermore we know the functions
$\frac{a_i}{a_j}$ for $0 \leq i \le j \leq N-1$ on $\P^{N-1}$ all
Poisson commute.  We could for example choose an algebraically
independent set of $N-1$ Hamiltonians $H_i \equiv \frac{a_i}{a_0}$,
$i=1\dots N-1$, which we wish to find as functions of $x_1, \dots,
x_{N-1}$.  We do this by solving the relation between $x_1, \dots,
x_{N-1}$ and $a_0, \dots a_{N-1}$ given by

\be\label{HamRel}
\left(\begin{array}{ccc}
\vth_0(x_1)      & \dots     & \vth_{N-1}(x_1)    \\
\vdots             & \ddots    & \vdots               \\
\vth_0(x_{N-1})  & \dots     & \vth_{N-1}(x_{N-1}) \\
\end{array}\right)
\left(\begin{array}{c}
a_0 \\
\vdots \\
a_{N-1} \\
\end{array}\right) = 0. \\
\ee
Let $\Theta$ be the $(N-1)\times N$ matrix appearing in \eqref{HamRel}.  Let
$\Theta[i]$ be the $(N-1)\times (N-1)$ matrix obtained from $\Theta$
by deleting
the $i$th column. Then the reader may check that solving for the $a_i$ in terms
of the $x_i$ in \eqref{HamRel} yields the desired Hamiltonians

\be\label{Hamiltonians}
\frac{a_i}{a_j} = (-1)^{i+j}\frac
          {\mathrm{Det}\,\Theta[i]}{\mathrm{Det}\,\Theta[j]}.
\ee
These Hamiltonians are manifestly symmetric in $x_1, \dots, x_{N-1}$ and
all Poisson commute with respect to the natural symplectic structure
on \Symm{^{N-1}T^4} given by
\be\label{eq:sstruct}
\omega = \sum_{i=1}^{N-1} du_1^{(i)} \wedge du_2^{(i)}.
\ee

\par It is instructive to consider the physics of the simplest case of
$SU(2)$ gauge theory, where $N$ is now $2$.  $\Msw$ is just $\P^1$
with homogenous coordinates $[a_0, a_1]$.  The point $[a_0, a_1] \in
\P^1$ corresponds to the genus 3 curve $a_0\vth_0 + a_1 \vth_1 =
0$.  The phase space $\cal M$ can be viewed as a $T^2$ fibration over
$\P^1$, or alternatively, away from nongeneric points, as a $T^4$ with
symplectic structure $\omega$ in \eqref{eq:sstruct}. The Hamiltonian
of the integrable system is $H = \frac{a_1}{a_0}$, or in terms of the
single 
separated variable $x$, $H = -\frac{\vth_0(x)}{\vth_1(x)}$.  The
Hamiltonian vector field $\chi_H$ is given as usual by the relation
$\omega (\chi_H, \centerdot ) = \partial H$.  It can be easily seen
that the equilibrium condition on $x = (u_1, u_2)$ with respect to
this Hamiltonian flow $\chi_H$ is equivalent to the two equations
\be\label{EqConfig}
\vth_1(x) \partial_{u_i} \vth_0(x) 
- \vth_0(x) \partial_{u_i} \vth_1(x) = 0 \qquad i=1,2.
\ee
These can be solved in two qualitatively distinct cases.
One is when $\vth_0(x) = \vth_1(x) = 0$.  This case actually occurs
at 4 points on $T^4$ as can be calculated from intersection theory.
The second case is nicely interpreted in terms of the
unique Seiberg-Witten curve that goes through $x$.  In terms of the moduli of
this curve, $[a_0, a_1]$, the equilibrium equations \eqref{EqConfig} can
be rewritten as
\be
a_0 \partial_{u_i} \vth_0(x) + a_1 \partial_{u_i} \vth_1(x) = 0 \qquad i=1,2.
\ee
However these are simply the conditions that the curve $a_0\vth_0 +
a_1 \vth_1$
is singular.  Thus we recover very easily the observations in lower dimensions
that equilibrium conditions on the phase space of the integrable system are
equivalent to degeneration conditions on the Seiberg-Witten curve, at least
for the case of $N=2$.

The proof for larger $N$ is suggested by the
above technique.  Recall the equation defining the Seiberg-Witten
curve:
\be
\sum_{i=0}^{N-1} a_i \vth_i(x_j) = 0
\ee
Now, as stated, this equation only holds for the specific $x_j$
defining the curve.  However, as we did above, one can solve for the
$a_i$ in terms of the $x_j$.  Defining $\vec x = (x_1,\dots,x_{N-1})$,
we see that the equation
\be
\sum_{i=0}^{N-1} a_i(\vec x) \vth_i(x_j) = 0
\ee
holds for any choice of $x_j$, or more generally for any choice of
$\vec x$.  Thus, we may take the derivative.  Letting
$x_j=(u_1^j,u_2^j)$, we see
\be\label{SWcurvederiv}
\sum_{i=0}^{N-1} \left[\left(\partial_{u_k^j}a_i(\vec x)\right) \vth_i(x_j)
  + a_i(\vec x) \partial_{u_k}\vth_i(x_j)\right] = 0;\quad
    k=1,2; \quad j=1,\dots, N-1.
\ee
The latter term is exactly that which appears in the condition for the
Seiberg-Witten curve to be degenerate.  If we show the first term is
0, then it follows that the curve degenerates.  To show this, use the
stationary-point condition for $H_i=a_i/a_0$.
\be
0  = \partial_{u_j^k}H_i
   = \frac {a_0\partial_{u_j^k}a_i - a_i\partial_{u_j^k}a_0}{a_0^2}
\ee
Now, assuming $a_0\ne0$ (otherwise we should have chosen a different
basis for our Hamiltonians), this is equivalent to
\be
\partial_{u_j^k}a_i = \frac{a_i}{a_0}\partial_{u_j^k}a_0.
\ee
Armed with this result, we can rearrange the first term above:
\bear
\sum_{i=0}^{N-1} \left(\partial_{u_k^j}a_i(\vec x)\right) \vth_i(x_j)
  &=& \sum_{i=0}^{N-1}
   \left(\frac{a_i}{a_0}\partial_{u_k^j}a_0(\vec x)\right) \vth_i(x_j)\\
  &=& \frac{1}{a_0}\partial_{u_k^j} a_0 \sum_{i=0}^{N-1}
        a_i(\vec x) \vth_i(x_j).
\eear
But this last sum is 0 because $x_j$ is on the curve.  Thus the second
term in Eq. (\ref{SWcurvederiv}) is similarly 0, and since this is
true for all $\partial_{u_j^k}$, the curve degenerates. 

The moral of the
story is that if a configuration of points $x_1,\dots,x_{N-1}$ is at
equilibrium with respect to all $N-1$ Hamiltonians, then the Seiberg-Witten
curve that goes through these points is singular at each and every one of them.
One can view this type of degeneration as $N-1$ pinched cycles.  Each time
a cycle pinches the genus is reduced by one.  Hence the generic genus $N+1$
Seiberg-Witten curve, at such special equilibrium points, has $N-1$ pinched
cycles and thus degenerates to genus 2.  This degeneration signals that
we are in a massive vacuum with no unbroken $U(1)$'s.  Furthermore this
degenerated curve should be related to the genus two spectral curve coming
from the matrix model.
 
\section{A Dual set of Hamiltonians}\label{sec:DualHams}

Here, inspired by \cite{MirMor:1999, Marshakov:1999}, we conjecture
the existence of an alternate set of Poisson commuting Hamiltonians
for the integrable system underlying the six dimensional $SU(N)$ gauge theory
which could shed some light on the physical nature of the system.  We begin
again with the genus $N+1$ Seiberg-Witten curves of our compactified \SUSY{2}
theory.  Their period matrices all exhibit an interesting structure, which 
is purely a consequence of the fact that they can all be embedded in a $T^4$.

\par Let $\Sigma_{N+1}$ be such a genus $N+1$ curve equipped with an embedding map
$\iota : \Sigma_{N+1} \rightarrow T^4$, with $T^4$ having a complex structure
specified by the period matrix \eqref{T4periods}.  Let $\alpha_1,\dots,\alpha_{N+1},
\beta_1,\dots,\beta_{N+1}$ be a symplectic basis of 1-cycles on $\Sigma_{N+1}$,
and $A_1,B_1,A_2,B_2$ be such a basis on $T^4$.  $\iota$ induces a map on 1-cycles
preserving the intersection product.  For convenience, we assume 
$\alpha_1,\beta_1 \rightarrow A_1, B_1$ under $\iota$ and all the rest map
to $A_2, B_2$.  The argument can be easily modified to accommodate a more general
situation.  Now $\Sigma_{N+1}$ has $N+1$ holomorphic one-forms.  Choose a basis
$\omega_1,\dots,\omega_{N+1}$ subject to the conditions
\bear
\omega_1 &= \iota^* \frac{du_1}{N} \\
\omega_2 + \cdots + \omega_{N+1} &= \iota^* du_2.
\eear
With this basis of one-forms, one can easily check that their periods over
the $\alpha$ cycles yield the identity matrix, and their periods over the
$\beta$ cycles yield the 
\be
\tilde T =
\left[\begin{array}{cccc}
\tau/N   & \alpha/N & \cdots & \alpha/N \\
\alpha/N & T_{11}   & \cdots & T_{1N} \\
\vdots   & \vdots   & \ddots & \vdots \\
\alpha/N & T_{N1}   & \cdots & T_{NN} \\
\end{array}\right],
\ee where $\sum_i T_{ij} = \sum_j T_{ij} = \rho$.  The motivation for
putting the period matrix in this form is to allow us to decompose 
a theta function on $\Jac{\Sigma_{N+1}} \sim T^{2N+2}$ into a sum over
products of theta functions on $T^4$ and theta functions on
$T^{2N-2}$.  Ratios of these latter theta functions should yield an
alternate set of Poisson commuting Hamiltonians.

\par We do this as follows.  Consider a theta function on $T^{2N+2}$,
$$
\vth(\tilde p|\tilde T) = \sum_{\tilde n \in \Z^{N+1}}
  \exp\left[
       \pi i \tilde n^t \tilde T \tilde n
       + 2\pi i \tilde n^t \tilde p
      \right].
$$
Decompose $\tilde n = (m_1, n)$ and $\tilde p = (q_1, p)$, where $m_1,
q_1$ are one-dimensional, while $n,p$ are $N$ dimensional.  The
curious subscripts anticipate the remainder of our decomposition.
With these, $\vth$ can be written as
$$
\vth(q_1,p|\tilde T) = \sum_{n\in\Z^N,m_1\in\Z}
  \exp\left[
       \pi i (m_1^2 \tau/N + 2 m_1 (\sum_i n) \alpha/N
       +n^t T  n) + 2\pi i (m_1 q_1 + n^t p)
      \right].
$$
Following the procedure laid down by \cite{MirMor:1999} and
\cite{Marshakov:1999}, let $T_{ij}=\rho/N + \hat T_{ij}$ and 
$p_i = q_2 + \hat p_i$, where
\be\label{hatsum}
\sum_i \hat p_i = \sum_i \hat T_{ij} = \sum_j \hat T_{ij} = 0.
\ee
Then, combining $q=(q_1,q_2)$, $m=(m_1,m_2)$, and $\Omega=
\left[\begin{array}{cc}
\tau & \alpha \\
\alpha & \rho \\
\end{array}\right]$,
our function becomes

\be
\vth(q,\hat p|\Omega,\hat T)
  = \sum_{m\in\Z^2}
    \exp\left[
         \pi i m^t \frac{\Omega}{N} m
         + 2\pi i m^t q
        \right]
      \sum_{ {n\in\Z^N} \atop {\Sigma_i n_i=m_2} }
    \exp\left[
         \pi i n^t \hat T n + 2\pi i n^t \hat p
        \right]
\ee
We express $m$ as $Nm + k$, where $k=0\dots N-1$.  This allows us to
sum over $\thx{\frac{k}{N}}{0}$
on $T^4$ (see Appendix \ref{sec:ThetaApx} for definitions).
\bear
\vth(q,\hat p|\Omega,\hat T)
  &=& \sum_{k\in\Z_N^2}
      \thx{\frac{k}{N}}{0}(Nq|N\Omega)
       \,\hat\vth^{(N-1)}_{k_2}(\hat p|\hat T)\mathrm{,\quad where}\\
\hat\vth^{(N-1)}_{k_2}(\hat p|\hat T)\label{eq:dualH}
  &=& \sum_{ {n\in\Z^N} \atop {\Sigma_i n_i = k_2} }
   \exp\left[
        \pi i n^t \hat Tn + 2\pi i n^t\hat p
       \right], \qquad k_2 = 0 \dots N-1.
\eear

In the limit where this integrable system degenerates to the elliptic
Calogero-Moser system relevant for the \SUSY{1^*} theory, ratios of the
$N$ functions defined in \eqref{eq:dualH} were shown to Poisson commute
in \cite{Marshakov:1999}.  It is reasonable to conjecture this continues
to holds true away from that limit.  Moreover, these functions where shown 
to be dependent only on the coordinates of the Calogero-Moser particles.  Hence
a deeper understanding of these functions might shed light on a 
particle/coordinate interpretation of the integrable system relevant for $6D$ 
gauge theory rather than just a phase space interpretation.

\par
Intuitively,  $\hat\vth^{(N-1)}_{k_2}(\hat p|\hat T)$ in \eqref{eq:dualH}
is a function on the part of $\Jac{\Sigma_{N+1}}$ that does not come
from $T^4$, and hence is a function of the angle variables.  However
$\hat\vth^{(N-1)}_{k_2}(\hat p|\hat T)$ also depends on the action 
variables though its dependence on $\hat T$.  If one thought of this
function as a function of the separated variables $x_1, \dots x_{N-1}$, 
then unlike the action variables, which depend only on the curve determined
by these $N-1$ points, $\hat\vth^{(N-1)}_{k_2}(\hat p|\hat T)$ also depends
on where these points are on the curve.  This particular combination
of action and angle variables could be interpreted as functions
of the ``coordinates'' of the integrable system.

\section{Summary and discussion}\label{sec:discussion}
In this paper we have made some progress towards exploring the circle
of ideas relating matrix models, Seiberg-Witten curves and integrable
systems in the context of massive vacua of twisted, compactified 6D
gauge theories.  We have observed a great deal of unity in the three
approaches, each of which ultimately gives its answer in the form of a
Riemann surface.  

\par In the matrix model approach the Riemann surface arises
microscopically through the condensation of matrix model eigenvalues.
Because these eigenvalues live on a compact two torus, the spectral
curve is of genus two.  Upon extremization of the DV superpotential,
the complex structure of the Jacobian of this Riemann surface is
related to a physical $T^4$ in which the Seiberg-Witten curves live.

\par In the integrable systems approach, the moduli space of vacua
upon further compactification is shown to be a system of $N-1$ points
living on this same $T^4$.  The relevant data here are a set of
Poisson commuting Hamiltonians, and the condition for a massive vacuum
is that the configuration of the integrable system be at equilibrium
with respect to all Hamiltonians.  \par This integrable system
condition is shown to reproduce the Seiberg-Witten condition for a
massive vacuum, namely that the Seiberg-Witten curve degenerates to
genus 2, which again is the same as the matrix model spectral curve.
As a further confirmation of these ideas it would be interesting to
calculate the actual values of the \SUSY{1} superpotential in the
massive vacua.  This would relate residue calculations on the matrix
model spectral curve to the values of integrable system Hamiltonians
at equilibrium points.  

\par In 4D the Lax-pair formulation of the integrable system provides
a nice dictionary between \SUSY{1} superpotentials and integrable
system Hamiltonians \cite{BBDW:2003}.  It would be useful to have this
dictionary in six dimensions.  One could also easily generalize
this paper to quiver models with $k$ hypermultiplets.  This would lead
to a fairly simple spin generalization of the system of points on
$T^4$, generalizing the spin generalizations found in lower dimensions.
Also a particle like interpretation of the 6D integrable system would be
useful, and might be related to the ansatz for a dual set of Poisson 
commuting Hamiltonians in section \ref{sec:DualHams}.

\acknowledgments
It is a pleasure to thank Eric Gimon, Petr Horava, Radu Tatar and Uday
Varadarajan for helpful discussions.
This work was supported in part by the Director, Office of Science,
Office of High Energy and Nuclear Physics, of the U.S. Department of
Energy under Contract DE-AC03-76SF00098, and in part by
the NSF under grant PHY-0098840.

\disclaimer
This document was prepared as an account of work sponsored by the United States Government.
While this document is believed to contain correct information, neither the United States
Government nor any agency thereof, nor The Regents of the University of California,
nor any of their employees, makes any warranty, express or implied,
or assumes any legal responsibility for the accuracy, completeness,
or usefulness of any information, apparatus, product, or process disclosed,
or represents that its use would not infringe privately owned rights.
Reference herein to any specific commercial product, process, or service by its trade name,
trademark, manufacturer, or otherwise,
does not necessarily constitute or imply its endorsement,
recommendation, or favoring by the United States Government or any agency thereof,
or The Regents of the University of California.
The views and opinions of authors expressed herein do not necessarily
state or reflect those of the United States Government or any agency thereof,
or The Regents of the University of California. 

Ernest Orlando Lawrence Berkeley National Laboratory is an equal opportunity employer.

\newpage
\appendix

\section{Meromorphic functions on $T^2$}\label{sec:elliptic}
Consider a torus $T^2$ with complex structure $\tau$ (Im $\tau>0$).
A theta
function $\vth$ on the torus is a quasi-periodic function, with the
following periodicity conditions:
\bear
\thx{a}{b}(z+1|\tau)=e^{2\pi ia}\thx{a}{b}(z|\tau)\\
\thx{a}{b}(z+\tau|\tau)=e^{-\pi i\tau-2\pi i (z+b)}\thx{a}{b}(z|\tau)
\eear
An explicit formula for $\vartheta$ is
\be
\thx{a}{b}(z|\tau) = \sum_{n=-\infty}^{\infty}
  \exp[\pi i (n+a)^2\tau + 2\pi i (n+a)(z+b)].
\ee
We can use three methods to construct meromorphic functions on $T^2$
from theta functions.  The first is to form ratios.
$$
\frac{\prod_{i=1}^{n} \thx{a_i}{b_i}( z|\tau)}
     {\prod_{i=1}^{n} \thx{a_i'}{b_i'}(z|\tau)}
$$
is a meromorphic function on $T^2$ provided
$\sum a_i \equiv \sum a_i',
 \sum b_i \equiv \sum b_i' \mod \mathbb{Z}$.
Another method is the derivative of the logarithm of a ratio of theta
function.
$$
\px{z_i}\ln\frac{\thx{a}{b}(z|\tau)}
                {\thx{a'}{b'}(z|\tau)}
$$
The last is the second derivative of the logarithm of a theta
function.
$$
\px{z_i}\px{z_j}\ln\thx{a}{b}(z|\tau)
$$
Using the transformation properties above, these can be shown to be
periodic in both directions.

Another method of constructing meromorphic functions on a $T^2$
involves the Weierstrass $\wp$-function, and its derivative
$\wp'$.  We define it as
\be
\wp(z)=\frac{1}{z^2} + \sum_{\lambda\in\Lambda,\lambda\ne0}
  \left(\frac{1}{(z-\lambda)^2}-\frac{1}{\lambda^2}\right),
\ee
where $\Lambda=\Z+\tau\Z$ is the lattice defining the torus.  This is
even, (doubly) periodic in $\Lambda$, analytic on
$\C\backslash\Lambda$, and has a pole of order two at the points on
$\Lambda$. $\wp$ and $\wp'$ satisfy the differential equation
\be
\wp^2 = 4\wp^3 - g_2 \wp - g_3,
\ee
where $g_2$ and $g_3$ are constants determined by the lattice
$\Lambda$ (and therefore $\tau$).  Note that as $\wp$ is even and
doubly periodic, $\wp'$ is odd and doubly periodic.  It turns out that
any doubly periodic function $F$ can be written as
\be
F(z) = R_1(\wp) + \wp'R_2(\wp),
\ee
with $R_1$ and $R_2$ rational functions.  Morally one decomposes $F$ 
into odd and even parts.

Two other Weierstrass functions deserve mention: the Weierstrass
$\sigma$ function and the Weierstrass $\zeta$ function, the latter not
to be confused with the Riemann $\zeta$ function.
We define them as
\bear
\sigma(z) &=& z\prod_{\lambda\in\Lambda,\lambda\ne0}
   \left(1-\frac{z}{\lambda}\right)
     \exp\left[\frac{z}{w} +
     \frac{1}{2}\left(\frac{z}{\lambda}\right)^2\right]\\
\zeta(z) &=& \frac{1}{z} + \sum_{\lambda\in\Lambda,\lambda\ne0}
   \left(\frac{1}{z-\lambda} + \frac{1}{\lambda}
     + \frac{z}{\lambda^2}\right).
\eear
$\zeta$ has a simple pole with residue 1 at every point in $\Lambda$,
and is analytic on $\C\backslash\Lambda$.
Lastly, we note the relations between these various functions and
their periodicity properties.
\bear
\zeta(z) &=& \frac{d}{dz}\log \sigma(z) \\
\wp(z) &=& -\frac{d}{dz}\zeta(z) \\
\zeta(z+n+m\tau) &=& \zeta(z) + n\eta_1 + m\eta_2 \\
\sigma(z +n+m\tau) &=& (-1)^{nm+n+m}\sigma(z)
         \exp\left[(n\eta_1 + m\eta_2)(z+\frac{1}{2}(n+m\tau))\right]\\
\eta_1\tau - \eta_2 &=& 2\pi i.
\eear

\section{Higher dimensional Theta functions}\label{sec:ThetaApx}
For a higher dimensional complex torus $\mathbb{C}^g/\Lambda$, where
$\Lambda$ is a lattice of rank $2g$, the analogy of 
the complex structure $\tau$ is a
$g\times g$ complex matrix $\Omega$.  $\Omega$ must be symmetric, and
$\mathrm{Im}\,\Omega$ must be positive definite.  Then the higher
dimensional $\vth$ functions are defined on $\mathbb{C}^g$ as
\be
\thx{\vec a}{\vec b}(\vec z|\Omega)=
  \sum_{\vec n\in\Lambda}\exp[\pi i
  (\vec n + \vec a)\cdot\Omega\cdot(\vec n+ \vec a)
  +2\pi i (\vec n + \vec a)\cdot(\vec z + \vec b)].
\ee
Similar to the $T^2$ case, where we could holomorphically transform
the lattice to $\mathbb{Z}+\tau\mathbb{Z}$, in the $g$-complex
dimensional case, we can view the lattice as $\Lambda =
\mathbb{Z}^g + \Omega\mathbb{Z}^g$.  The periodicity properties are
also analogous.  Let $\vec m \in \mathbb{Z}^g$.
\bear
\thx{\vec a}{\vec b}(\vec z + \vec m|\Omega)
  &=& e^{2\pi i \vec a\cdot\vec m}\thx{\vec a}{\vec b}(\vec z|\Omega),\\
\thx{\vec a}{\vec b}(\vec z + \Omega\vec m|\Omega)
  &=& e^{-\pi i \vec m\cdot\Omega\vec m-2\pi i \vec m\cdot(\vec z + \vec b)}
   \thx{\vec a}{\vec b}(\vec z|\Omega).
\eear

%

We can use the same three methods as before to construct meromorphic
functions on
$T^{2g}=\mathbb{C}^g/\Lambda$ using theta functions.  We repeat them
here in the multi-dimensional notation for appendectical completeness.
\be
\frac{\prod_{i=1}^{n} \thx{\vec a_i}{\vec b_i}(\vec z|\Omega)}
     {\prod_{i=1}^{n} \thx{\vec a_i\prime}{\vec b_i\prime}(\vec
       z|\Omega)}
\ee
is a meromorphic function on $T^{2g}$ provided
$\sum a_i \equiv \sum a_i',
 \sum b_i \equiv \sum b_i' \mod \mathbb{Z}^g$.

\be \mathrm{So\ are}\quad
\px{z_i}\ln\frac{\thx{\vec a}{\vec b}(\vec z|\Omega)}
                {\thx{\vec a'}{\vec b'}(\vec z|\Omega)}
\quad \mathrm{and} \quad
\px{z_i}\px{z_j}\ln\thx{\vec a}{\vec b}(\vec z|\Omega),
\ee
for any choice of characters.  Again, using the above transformation
properties, these can be shown to be periodic in all $2g$ directions.

These functions can be used to define functions on genus $g$
Riemannian surfaces.  Consider such a $\Sigma_g$.  There are $g$
holomorphic 1-forms on $\Sigma_g$, call them $\omega_i$.  Denote the
canonical basis of $H_1(\Sigma_g,\Z)$ by the $2g$ cycles $A_i$ and
$B_i$, where $A_i\cap A_j=B_i\cap B_j=0$, and $A_i\cap
B_j=\delta_{ij}$.  Then we can define the $g\times 2g$
\emph{period matrix} as
\be\label{TgPeriods}
\left[\begin{array}{cccccc}
\int_{A_1}\omega_1 & \cdots & \int_{A_g}\omega_1 &
 \int_{B_1}\omega_1 & \cdots & \int_{B_g}\omega_1 \\
\vdots & \quad & \vdots & \vdots & \quad & \vdots \\
\int_{A_1}\omega_g & \cdots & \int_{A_g}\omega_g &
 \int_{B_1}\omega_g & \cdots & \int_{B_g}\omega_g \\
\end{array}\right].
\ee
We can choose the $\omega_i$ such that
$\int_{A_j}\omega_i=\delta_{ij}$; then the period matrix is in the
form $[\Id,\Omega]$ for the
$g\times g$ identity matrix $\Id$ and a $g \times g$ symmetric matrix
$\Omega$, where Im $\Omega$ is positive definite.  This similarity
with the complex structure matrix in the beginning of this appendix,
which was cunningly also named $\Omega$, is not coincidental, as we
now show.

Now the columns of the period matrix are $2g$ vectors in $\C^g$; these
naturally form a lattice $\Lambda$ and thus induce a torus
$T^{2g}=\C^g/\Lambda$.
This torus is called the \emph{Jacobian} of $\Sigma_g$, often denoted
$\Jac{\Sigma_g}$.
What is the relation between these two objects?  The
answer is given by the Abel-Jacobi map.

Choose a $p_0 \in \Sigma_g$, and consider the function 
$\mu:\Sigma_g \rightarrow \C^g/\Lambda$, under which
\be
p \mapsto \left(\int_{p_0}^p \omega_1, \dots,
    \int_{p_0}^p \omega_g\right).
\ee
Note this is only defined up to $\Lambda$, since in choosing a contour
from $p_0$ to $p$ we could go around any combination of the cycles of
the torus.  In fact, we can generalize this as a map from any degree 0
divisor to $\mathcal J$.  This is the Abel-Jacobi map,
$\mu:\mathrm{Div^0 }\Sigma_g \rightarrow \Jac{\Sigma_g}$, where
\be
\sum_i(p_i - q_i) \mapsto \left(\sum_i \int_{q_i}^{p_i} \omega_1,
   \dots, \sum_i \int_{q_i}^{p_i} \omega_g\right).
\ee

As $\Sigma_g$ is one-complex dimensional, and $\mathcal J$ is
$g$-complex dimensional, in order to get a surjective map we need to
pick $g$ points on $\Sigma_g$, say $p_i$, $i=1,\dots,g$.  The Jacobi
Inversion Theorem states this explicitly:  Given any
$\lambda \in \Jac{\Sigma_g}$, there exist $g$ points
$p_i\in\Sigma_g$ such that $\mu\left(\sum_i(p_i-p_0)\right)=\lambda$.
Moreover, these $p_i$ are generically unique.  Finally, Abel's theorem
states that if the divisor $\sum(p_i-q_i)$ is a divisor of some
meromorphic function, then $\mu\left(\sum(p_i-q_i)\right)=0$.  These two
results mean the Abel-Jacobi map is an isomorphism between the moduli
space of line bundles of degree 0, $\mathrm{Pic}^0(\Sigma_g)$, and the
Jacobian $\Jac{\Sigma_g}$.

As promised, this allows us to find meromorphic functions on
$\Sigma_g$.  The Abel-Jacobi map gives us an embedding of $\Sigma_g$ into
\Jac{\Sigma_g}.  By constructing meromorphic functions on
$T^{2g}\cong \Jac{\Sigma_g}$, we can simply pull them back under the
Abel-Jacobi map to get meromorphic functions on $\Sigma_g$.

More complete expositions can be found in \cite{GrHa} and \cite{Tata}.


\newpage 
\begin{thebibliography}{1}

\bibitem{Dijkgraaf:2002fc}
R.~Dijkgraaf and C.~Vafa,
{``Matrix models, topological strings, and supersymmetric gauge theories,''}
Nucl.\ Phys.\ B {\bf 644}, 3 (2002)
[arXiv:hep-th/0206255].

\bibitem{Dijkgraaf:2002vw}
R.~Dijkgraaf and C.~Vafa,
{``On geometry and matrix models,''}
Nucl.\ Phys.\ B {\bf 644}, 21 (2002)
[arXiv:hep-th/0207106].

\bibitem{Dijkgraaf:2002dh}
R.~Dijkgraaf and C.~Vafa,
{``A perturbative window into non-perturbative physics,''}
arXiv:hep-th/0208048.

\bibitem{Cachazo:2002ry}
F.~Cachazo, M.~R.~Douglas, N.~Seiberg and E.~Witten,
{``Chiral rings and anomalies in supersymmetric gauge theory,''}
JHEP {\bf 0212}, 071 (2002)
[arXiv:hep-th/0211170].

\bibitem{Cachazo:2002zk}
F.~Cachazo, N.~Seiberg and E.~Witten,
{``Phases of N = 1 supersymmetric gauge theories and matrices,''}
JHEP {\bf 0302}, 042 (2003)
[arXiv:hep-th/0301006].

\bibitem{Roiban:2003uq}
R.~Roiban, R.~Tatar and J.~Walcher,
{``Massless flavor in geometry and matrix models,''}
arXiv:hep-th/0301217.

\bibitem{Cachazo:2003yc}
F.~Cachazo, N.~Seiberg and E.~Witten,
{``Chiral Rings and Phases of Supersymmetric Gauge Theories,''}
JHEP {\bf 0304}, 018 (2003)
[arXiv:hep-th/0303207].

\bibitem{Petrini:2003py}
M.~Petrini, A.~Tomasiello and A.~Zaffaroni,
{``On the geometry of matrix models for N = 1,''}
JHEP {\bf 0308}, 004 (2003)
[arXiv:hep-th/0304251].

\bibitem{Bena:2003vk}
I.~Bena, H.~Murayama, R.~Roiban and R.~Tatar,
{``Matrix model description of baryonic deformations,''}
arXiv:hep-th/0303115.

\bibitem{Alishahiha:2003pu}
M.~Alishahiha and A.~E.~Mosaffa,
{``On effective superpotentials and compactification to three dimensions,''}
arXiv:hep-th/0304247.


\bibitem{Veneziano:1982ah}
G.~Veneziano and S.~Yankielowicz,
{``An Effective Lagrangian For The Pure N=1 Supersymmetric Yang-Mills Theory,''}
Phys.\ Lett.\ B {\bf 113}, 231 (1982).

\bibitem{Intriligator:1994jr}
K.~A.~Intriligator, R.~G.~Leigh and N.~Seiberg,
{``Exact superpotentials in four-dimensions,''}
Phys.\ Rev.\ D {\bf 50}, 1092 (1994)
[arXiv:hep-th/9403198].

\bibitem{Seiberg:1994bp}
N.~Seiberg,
{``The Power of holomorphy: Exact results in 4-D SUSY field theories,''}
arXiv:hep-th/9408013.

\bibitem{Seiberg:1995ac}
N.~Seiberg,
{``The power of duality: Exact results in 4D SUSY field theory,''}
Int.\ J.\ Mod.\ Phys.\ A {\bf 16}, 4365 (2001)
[arXiv:hep-th/9506077].

\bibitem{Seiberg:1994rs}
N.~Seiberg and E.~Witten,
{``Electric - magnetic duality, monopole condensation, and confinement in
 $N=2$ supersymmetric Yang-Mills theory,''}
Nucl.\ Phys.\ B {\bf 426}, 19 (1994)
[Erratum-ibid.\ B {\bf 430}, 485 (1994)]
[arXiv:hep-th/9407087].

\bibitem{Seiberg:1994aj}
N.~Seiberg and E.~Witten,
{``Monopoles, duality and chiral symmetry breaking in N=2 supersymmetric QCD,''}
Nucl.\ Phys.\ B {\bf 431}, 484 (1994)
[arXiv:hep-th/9408099].





\bibitem{Dijkgraaf:2003xk}
R.~Dijkgraaf and C.~Vafa,
{``$N = 1$ supersymmetry, deconstruction, and bosonic gauge theories,''}
arXiv:hep-th/0302011.

\bibitem{Hollowood:2003gr}
T.~J.~Hollowood,
{``Five-dimensional gauge theories and quantum mechanical matrix models,''}
JHEP {\bf 0303}, 039 (2003)
[arXiv:hep-th/0302165].


\bibitem{Cheung:1998te}
Y.~K.~Cheung, O.~J.~Ganor and M.~Krogh,
{``On the twisted (2,0) and little-string theories,''}
Nucl.\ Phys.\ B {\bf 536}, 175 (1998)
[arXiv:hep-th/9805045].


\bibitem{Cheung:1998wj}
Y.~K.~Cheung, O.~J.~Ganor, M.~Krogh and A.~Y.~Mikhailov,
{``Instantons on a non-commutative $T^4$
from twisted $(2,0)$ and  little-string theories,''}
Nucl.\ Phys.\ B {\bf 564}, 259 (2000)
[arXiv:hep-th/9812172].


\bibitem{Chan:2000qc}
C.~S.~Chan, O.~J.~Ganor and M.~Krogh,
{``Chiral compactifications of 6D conformal theories,''}
Nucl.\ Phys.\ B {\bf 597}, 228 (2001)
[arXiv:hep-th/0002097].

\bibitem{Hollowood:2003cv}
T.~J.~Hollowood, A.~Iqbal and C.~Vafa,
{``Matrix Models, Geometric Engineering and Elliptic Genera,''}
arXiv:hep-th/0310272.


\bibitem{Braden:2003}
H.~W.~Braden and T.~J.~Hollowood
{``Critical Points of Glueball Superpotentials and Equilibria of Integrable Systems''}
[arXiv:hep-th/0311024].

\bibitem{Tata}
D.~Mumford,
{\it Tata Lectures on Theta I and II,} Birkhauser, 1982.

\bibitem{Hollowood:2003cp}
T.~J.~Hollowood
{``Critical Points of Glueball Superpotentials and Equilibria of Integrable Systems''}
[arXiv:hep-th/0305023].


\bibitem{DonagiWitten:1995}
R.~Donagi and E.~Witten,
{``Supersymmetric Yang-Mills Systems and Integrable Systems,''}
Nucl.\ Phys.\ B {\bf 460}, 299-334 (1996)
[arXiv:hep-th/9510101].

\bibitem{SeibergWitten:1996}
N.~Seiberg and E.~Witten,
{``Gauge Dynamics and Compactification To Three Dimensions,''}
[arXiv:hep-th/9607163].

\bibitem{Dorey:2002}
N.~Dorey, T.~Hollowood, S.~Prem Kumar and A.~Sinkovics,
{``Exact Superpotentials from Matrix Models''}
[arXiv:hep-th/0209089].

\bibitem{Dorey:1999}
N.~Dorey,
{``An Elliptic Superpotentional for Softly Broken \SUSY{4}
Supersymmetric Yang-Mills Theory''}
[arXiv:hep-th/9906011].

\bibitem{BBDW:2003}
R.~Boels, J.~de Boer, R.~Duivenvoorden and J. Wijnhout,
{``Nonperturbative Superpotentials and Compactification to Three Dimensions,''}
[arXiv:hep-th/0304061].

\bibitem{Ganor:2000un}
O.~J.~Ganor, A.~Y.~Mikhailov and N.~Saulina,
{``Constructions of non commutative instantons on $T^4$ and $K_3$,''}
Nucl.\ Phys.\ B {\bf 591}, 547 (2000)
[arXiv:hep-th/0007236].

\bibitem{Sklyanin:1995}
E.~K.~Sklyanin
{``Separation of Variables. New Trends''}
[arXiv:solv-int/9504001].

\bibitem{GNR:1999}
A.~Gorsky, N.~Nekrasov, V.~Rubstov,
{``Hilbert Schemes, Separated Variables, and D-Branes''}
[arXiv:hep-th/9901089].

\bibitem{MirMor:1999}
A.~Mironov and A.~Morozov,
{``Commuting Hamiltonians from Seiberg-Witten $\Theta$-Functions,''}
[arXiv:hep-th/9912088].

\bibitem{Marshakov:1999}
A.~Marshakov,
{``Duality in Integrable Systems and Generating Functions for New
  Hamiltonians,''}
[arXiv:hep-th/9912124].

\bibitem{GrHa}
P. Griffith and J. Harris,
{\it Introduction to algebraic geometry,} John Wiley \& Sons 1994.


\bibitem{Gorsky:1997mw}
A.~Gorsky, S.~Gukov and A.~Mironov,
{``SUSY field theories, integrable systems and their stringy/brane  origin.
II,''}
Nucl.\ Phys.\ B {\bf 518}, 689 (1998)
[arXiv:hep-th/9710239].

\bibitem{Nekrasov:1996cz}
N.~Nekrasov,
{``Five dimensional gauge theories and relativistic integrable systems,''}
Nucl.\ Phys.\ B {\bf 531}, 323 (1998)
[arXiv:hep-th/9609219].

\end{thebibliography}
\end{document}